\documentclass[sort&compress,AMA,STIX1COL]{WileyNJD-v2}
\usepackage{amsmath}
\articletype{RESEARCH ARTICLE}%

\received{ x 2020}
\revised{ x x}
\accepted{ x x}

\begin{document}

\title{Parameter clustering in Bayesian functional PCA of neuroscientific data}

\author[1]{Nicol\`{o} Margaritella*}

\author[1]{Vanda In\'{a}cio}

\author[1]{Ruth King}

\authormark{Nicol\`{o} Margaritella \textsc{et al}}

\address[]{\orgdiv{School of Mathematics}, \orgname{University of Edinburgh}, \orgaddress{\state{Edinburgh}, \country{UK}}}

\corres{*Nicol\`{o} Margaritella,\\
	University of Edinburgh,\\
	School of Mathematics,\\
	James Clerk Maxwell Building,\\
	The King's Buildings,\\
	Peter Guthrie Tait Road,\\
	Edinburgh,
	UK. \\ \email{N.Margaritella@sms.ed.ac.uk}}

\abstract[Summary]{The extraordinary advancements in neuroscientific technology for brain recordings over the last decades have led to increasingly complex spatio-temporal datasets. To reduce oversimplifications, new models have been developed to be able to identify meaningful patterns and new insights within a highly demanding data environment. To this extent, we propose a new model called parameter clustering functional Principal Component Analysis (PCl-fPCA) that merges ideas from Functional Data Analysis and Bayesian nonparametrics to obtain a flexible and computationally feasible signal reconstruction and exploration of spatio-temporal neuroscientific data. In particular, we use a Dirichlet process Gaussian mixture model to cluster functional principal component scores within the standard Bayesian functional PCA framework. This approach captures the spatial dependence structure among smoothed time series (curves) and its interaction with the time domain without imposing a prior spatial structure on the data. Moreover, by moving the mixture from data to functional principal component scores, we obtain a more general clustering procedure, thus allowing a higher level of intricate insight and understanding of the data. We present results from a simulation study showing improvements in curve and correlation reconstruction compared with different Bayesian and frequentist fPCA models and we apply our method to functional Magnetic Resonance Imaging and Electroencephalogram data analyses providing a rich exploration of the spatio-temporal dependence in brain time series.}


\jnlcitation{\cname{%
\author{N. Margaritella}, 
\author{V. In\'{a}cio}, and
\author{R. King}} (\cyear{2020}), 
\ctitle{Parameter clustering in Bayesian functional PCA of neuroscientific data}, \cjournal{Stat. Med.}, \cvol{2020;00:--}.}

\maketitle


\section{INTRODUCTION}\label{sec1}

Several tools for the recording of different brain processes, such as functional Magnetic Resonance Imaging (fMRI) and Electroencephalogram (EEG) produce remarkable amounts of spatio-temporal data which challenge researchers to find suitable models for increasingly complex datasets. Consequently, the last decade has seen a marked increase in the development flexible methods for high dimensional data in neuroscience.
Functional Data Analysis (FDA) is a fairly recent research field in statistics concerned with the analysis of data providing information about curves, shapes and images which vary over a continuum, usually time or space (see Ramsay and Silverman\cite{ramsay2005functional} for an overview). In the FDA framework, data can be considered as noise-corrupted, discretised realisations of underlying smooth functions (curves or trajectories) which are recovered using basis expansions and smoothing.\cite{ramsay2009functional} Many standard statistical tools have been translated into the FDA framework. Functional Principal Component Analysis (fPCA) is a technique that defines a set of smooth trajectories as an expansion of orthonormal bases (eigenfunctions) and weights which are called functional principal component scores (fPC scores).\cite{ramsay2005functional} One of the advantages of fPCA  is that it can be conveniently represented as a hierarchical mixed model in the Bayesian setting, with the joint posterior distribution of the fPC scores being the main target of inference.\cite{crainiceanu2010bayesian}

There has been a growing interest in applying FDA to neuroscientific data (see, among others, Viviani et al.,\cite{viviani2005functional} Tian et al.\cite{tian2010functional} and Hasenstab et al.\cite{hasenstab2017multi}). Often, in the FDA literature, underlying random curves are assumed to be independent and their correlation is ignored if believed to be mild.\cite{liu2017functional} However, curve dependence is of particular importance in the analysis of brain activity because of the complex architecture of spatio-temporal connections between brain areas.\cite{wolfson2018understanding}
Recently, Liu et al.\cite{liu2017functional} considered spatial dependence among trajectories by modelling the covariance of the fPC scores within a frequentist approach. Their results showed significant improvements in curve reconstruction compared to the standard approach assuming independence, especially with low signal-to-noise ratios.

The present study introduces a new method for the analysis of functional data in neuroscience. We develop a novel Bayesian fPCA model called Parameter Clustering fPCA (PCl-fPCA) that makes use of a Dirichlet Process (DP) mixture \cite{escobar1995bayesian,neal2000markov,rasmussen2000infinite} to model the prior distribution of the fPC scores. Different functional mixture models that cluster functions through clustering of the coefficients in a basis expansion have been proposed in the literature.\cite{james2003clustering,ray2006functional,zhou2006bayesian,dunson2008bayesian,bigelow2009bayesian,rodriguez2009bayesian,angelini2012clustering,scarpa2014enriched} However, these works have focused on a global clustering of curves, without considering local differences as well as the possibility of a dynamic evolution of dependence among curves. In this work we use the principal component bases due to their straightforward interpretation and employ DP mixture priors for every eigendimension retained. By allowing different clustering of the fPC scores for each eigendimension retained, we avoid the limitations of assuming separability of the cross-covariance and any a priori spatial covariance structure of the data, obtaining  further insights from space-time interactions. 

The study of how interactions among brain regions change dynamically during an experiment (i.e. dynamic functional connectivity) has recently attracted wide interest in the neuroimaging literature. This analysis has the potential to improve our understanding of how the brain works under both physiological and pathological conditions with recent studies focusing on the application of dynamic functional connectivity to aging,\cite{xia2019tracking} schizophrenia,\cite{dong2019reconfiguration} dementia and Parkinson's disease.\cite{fiorenzato2019dynamic} This is a new frontier for neuroscientific research and the development of suitable models able to capture the intricate spatio-temporal dynamics in the data will lay the foundations for the progress in this area in coming years.\cite{hutchison2013dynamic}

In this regard, we show that our approach has multiple advantages in the analysis of neuroscientific data as it offers further insights into the spatio-temporal structure of the data as a result of dimension-specific curve classification; it improves curve reconstruction thanks to the local borrowing of information compared to current fPCA approaches; and it can be defined as a simple and computationally feasible hierarchical model which can be easily implemented in \texttt{R}.  

The rest of the paper is structured as follows: in Section \ref{sec2} we overview the standard Bayesian fPCA model and introduce our new method, along with computational details. Section \ref{sec3} reports the setting and results of a simulation study where we compare the performance of PCl-fPCA with standard Bayesian and frequentist fPCA approaches under different data generating processes and noise levels. Section \ref{sec4} addresses the application of our method to a resting-state fMRI dataset and a task-based EEG recording and we discuss the further insights obtained in the spatio-temporal structure of the data and the underlying neurophysiological processes. Conclusions are discussed in Section \ref{sec5}.

\section{Methods}
\label{sec2}
\subsection{Bayesian Functional PCA}
\label{sec2.1}

The standard FDA model is given by
\begin{equation}\label{eq1}
Y_{it}= X_{it}+ \epsilon_{it},
\end{equation}
where $ Y_{it} $ denote the noise-corrupted, discretised, observed data for every spatially-correlated region (trajectory) $ i=1,\dots,n $ and time point $ t=1,\dots,T $; $ X_{it} $ the associated underlying random curve as a realisation of an $ L^{2} $ stochastic process $ \left\lbrace  X_{t} :t\in [1,T] \subseteq \mathcal{R}\right\rbrace  $ with mean $ \mu_{t} $ and covariance function $ G(s,t) $; and $\epsilon_{it}$ the noise term with zero mean and precision $ \tau $.\cite{wang2015review}

Functional PCA assumes that the covariance kernel $ G(s,t) $ 
of the process $ X_{t} $ can be represented by the Karhunen-Lo\`{e}ve expansion, such that
\begin{align}\label{eq2}
G(s,t)&=\sum_{k=1}^{\infty}\lambda_{k}\phi_{kt}\phi_{ks},\quad s,t \in [1,T],\\
X_{it}&= \mu_{t}+\sum_{k=1}^{\infty}\xi_{ik}\phi_{kt},\quad i=1,\ldots,n,
\end{align}
where $ \phi_{kt} $ are orthonormal eigenfunctions and $ \lambda_{k} $ are the associated eigenvalues. Then, each realisation $ X_{it} $ can be represented by a linear combination of eigenfunctions $ \phi_{kt} $, which are usually assumed to be observed, and fPC scores $ \xi_{ik} $, which are the main goal of inference. The reader is referred to Chapter 8 of Ramsay and Silverman \cite{ramsay2005functional} and the recent review of Joliffe and Cadima \cite{jolliffe2016principal} for a more detailed presentation of functional PCA. Although the number of eigendimensions can also be modelled with an appropriate distribution (see, for example, Suarez et al. \cite{suarez2017bayesian}), this considerably increases the computational complexity of the model and thus in practice only $ K $ pre-determined terms of the linear expansion are retained pertaining to those that explain a sufficiently large part of the total variability in the data.\cite{sorensen2013introduction} Often the case $\mu_{t}=0 $ is assumed and the centred data $ \tilde{Y}_{it} $ are obtained by subtracting an estimate $\widehat{\mu}_{t}$ of the population average. \cite{crainiceanu2010bayesian} 

The fPC scores $ \xi_{ik} $ are given prior probability distributions in the Bayesian framework. The standard Bayesian fPCA model \cite{crainiceanu2010bayesian} assumes fPC scores to be independent draws from a univariate zero-centred normal distribution whose variance is dependent on the eigendimension $ k $. The most straightforward hierarchical representation of the standard Bayesian fPCA model is
\begin{eqnarray}\label{eq3}
\tilde Y_{it}&= &\sum_{k=1}^{K}\xi_{ik}\phi_{kt} + \epsilon_{it},\\\nonumber 
\xi_{ik}|s_{k}&\sim &\text{N}(0,s^{-1}_{k}),\\\nonumber
\epsilon_{it}|\tau&\sim &\text{N}(0,\tau^{-1}),\\\nonumber
s_{k} &\sim &\Gamma(a,b),\\\nonumber
\tau &\sim &\Gamma(a',b'),
\end{eqnarray}
with $ a, a', b, b' $ usually set to low values (e.g. $ 10^{-3} $). In this model the noise term is assumed to be Gaussian and independent gamma priors are placed over the precision parameters because of their conjugacy property, permitting closed-form conditional posterior distributions and the use of Gibbs sampling.

Recently, Liu et al.\cite{liu2017functional} proposed to capture spatial dependence through a suitable model for the covariance of fPC scores. In particular, they defined $ \text{Cov}(\xi_{ik},\xi_{i'k}) $ as a function of the correlation coefficient $ \rho_{ii'k} $ which they modelled using the Mat\'{e}rn function family and estimated the corresponding parameters. This approach implies the a priori definition of a covariance structure which depends on the distance between observations; such assumptions might not be suitable for complex spatio-temporal phenomena such as brain activity where dependencies are the result of both structural and functional neuronal pathways as well as task-specific characteristics.
In this study, we overcome these limitations to achieve a higher level of flexibility in the modelling of the spatio-temporal covariance of neuroscientific data.

\subsection{PCl-fPCA model}
\label{sec2.2} 

In this section we present the structure of the PCl-fPCA model and the features of this approach that improve the current methods for functional PCA. 
The following hierarchical model defines the probability distribution generating observed time series. We present and comment on each level separately.

\textit{Level 1:} As the standard Bayesian fPCA model in Equation (\ref{eq3}), the distribution of the centred data given the parameters of the underlying smooth function and the noise term is given by:
\begin{align}\label{eq4}
\tilde{\textbf{Y}}_{i}|\textbf{X}_{i},\tau &\sim \text{N}_{\text{T}}(\textbf{X}_{i},\tau^{-1}\textbf{I}),\\\nonumber
\textbf{X}_{i}&=\sum_{k=1}^{K}\xi_{ik}\boldsymbol{\phi}_{k},
\end{align}
where $ \tilde{\textbf{Y}}_{i} $, $ \textbf{X}_{i} $ and $ \boldsymbol\phi_{k} $ are \textit{T}-dimensional vectors and $ \text{N}_{\text{T}}(\textbf{X}_{i},\tau^{-1}\textbf{I})  $ denotes a multivariate Gaussian distribution with mean $ \textbf{X}_{i} $ and variance-covariance matrix $ \tau^{-1}\textbf{I} $ such that $ \textbf{I} $ denotes the $ T\times T $ identity matrix. As in Equation (\ref{eq3}), the eigenfuctions $ \boldsymbol\phi_{k} $ are assumed to be observed and the parameter $ \tau $ does not depend on $ i $ or $ t $, i.e. the noise is assumed to be constant in both space and time, although other characterisations are possible.\cite{wang2015review} It follows that the likelihood function is given by
\begin{align}\label{eq5}
\text{L}(\tilde{\textbf{Y}}|\textbf{X},\tau)&= \Big(\dfrac{\tau}{2\pi}\Big)^{Tn/2}\exp\Big\{-\frac{\tau}{2}\sum_{i=1}^{n}(\tilde{\textbf{Y}}_{i}-\textbf{X}_{i})^{'}(\tilde{\textbf{Y}}_{i}-\textbf{X}_{i})\Big\}.
\end{align}

\textit{Level 2:} To encode fPC scores cluster membership we introduce a classification variable $ c_{ik} $ as a stochastic indicator that identifies which latent class $ j $ in eigendimension $ k $ is associated with parameter $ \xi_{ik} $. Prior distributions of the fPC scores $ \xi_{ik} $, given the parameters of underlying clusters $ \big[(\mu_{1k},s_{1k}),\dots,(\mu_{Jk},s_{Jk})\big] $ and the classification variable $ c_{ik} $, are given by
\begin{equation}\label{eq6}
\xi_{ik}|c_{ik},\mu_{1k},\dots,\mu_{Jk},s_{1k},\dots,s_{Jk}\sim\text{N}\big(\mu_{c_{ik}},s^{-1}_{c_{ik}}\big),
\end{equation}
where $ \mu_{c_{ik}=j} $ and $ s_{c_{ik}=j} $ denote the mean and precision for the $ j $-th cluster in the $ k $-th eigendimension, respectively. Here we use a $ J- $dimensional mixture of Gaussian distributions, independently, for each retained eigendimension $ k=1,\dots,K $ as we permit different (independent) partitions of the fPC scores for each mode of variation. It is worth recalling that, in the context of DP mixtures, $ J $ represents an upper-bound on the number of fPC score clusters.\cite{dunson2010nonparametric} In the rest of the manuscript we define  $ J_{k}^{+} < J $ as the (data-driven) number of non-empty clusters in each eigendimension $ k $.\cite{griffiths2005} 

\textit{Level 3:} Prior distributions for $ \big[(\mu_{1k},s_{1k}),\dots,(\mu_{Jk},s_{Jk})\big]  $ and $ (c_{1k},\dots,c_{nk}) $, given hyperparameters $ r_{k},\,\beta_{k} $ and parameters $ (p_{1k},\dots,p_{Jk}) $, are given by
\begin{align}\label{eq7}
c_{1k},\dots,c_{nk}|p_{1k},\dots,p_{Jk}&\sim f_{\text{C}}\big(p_{1k},\dots,p_{Jk}\big),\\\nonumber
\mu_{jk}|r&\sim \text{N}(0,r^{-1}_{k}),\\\nonumber
s_{jk}|\beta&\sim \Gamma(1,\beta_{k}),
\end{align}
where $f_{\text{C}} $ denotes the categorical distribution which generalises the Bernoulli random variable to $ J $ outcomes. Cluster precision $ s_{jk} $ can also be modelled using Uniform distributions on the cluster standard deviation where $ \sigma_{jk}=1/\sqrt{(s_{jk})} $.\cite{gelman2006prior}
Hyperparameters $ r$ and $\beta $ are often centred around empirical estimates in the literature \cite{richardson1997bayesian}; here, we take advantage of the properties of fPCA decomposition to tune the higher hierarchical levels in our model around weakly informative prior distributions.
It follows from the Karhunen-Lo\`{e}ve representation that, for any given $ i $, $ \xi_{ik} $ are uncorrelated fPC scores with monotonically decreasing variance given by the eigenvalues $ \lambda_{k} $ \cite{liu2017functional}; therefore, sensible functions of the empirical estimates of the eigenvalues $ \widehat{\lambda}_{k} $ can be used to fix $ r$ and $\beta $ under the assumption that, for every eigendimension $ k $, the position and dispersion of a cluster are both functions of $ \widehat{\lambda}_{k} $. We note that  setting $r= 1/\widehat{\lambda}_{k} $ and $\beta=\widehat{\lambda}_{k} $ worked well in our simulations and application.

\textit{Level 4 and 5:} Prior distribution for $ (p_{1k},\dots,p_{Jk}) $, given hyperparameter $ \alpha $ and prior distribution for $ \alpha $ are given by
\begin{align}\label{eq8}
p^{'}_{jk}|\alpha_{k}&\sim\text{Beta}(1,\alpha_{k}),\\\nonumber
p_{1k}=\dfrac{p^{'}_{1k}}{\sum_{j=1}^{J}p^{'}_{jk}};&\quad
p_{jk}=\dfrac{p^{'}_{jk}\prod_{l<j}(1-p_{lk})}{\sum_{j=1}^{J}p^{'}_{jk}},\quad j=1,\dots,J\\\nonumber
\alpha_{k}&\sim \text{U}[0,Q_{k}],
\end{align}
where $p_{jk}$ follow the stick-breaking construction \cite{sethuraman1994constructive} with parameter $ \alpha_{k} $ modelling the prior belief over the mixing proportions $ p_{1k},\dots,p_{Jk} $. The dispersion parameter $ \alpha $ is usually fixed or modelled with a prior distribution; here we used a uniform distribution with sufficiently large $ Q $.\cite{rasmussen2000infinite,medvedovic2002bayesian,medvedovic2004bayesian,de2018bayesian}\\

Different specifications of $ s_{jk} $ and $ Q $ can be employed for $ k=1 $ and $ k=2,\dots,K $ to incorporate the knowledge that the first eigendimension is more likely to capture global patterns in the data while the following dimensions are more sensitive to local features. For example, in the first eigendimension one can use the gamma distribution for the cluster precision in Equation~(\ref{eq7}) as it assigns more weights to large clusters than a uniform on the standard deviation which can be used instead in the subsequent dimensions. We provide specific examples in Section \ref{sec3.1} and the results of a sensitivity analysis on $ Q,\beta$ and $ s $ in the WebA section of the Supplementary Material file.

The model structure can be displayed with a direct acyclic graph (DAG) (Supplementary Material, WebB section, Figure A1).
As $ J $ approaches infinity the model corresponds to a DP mixture model\cite{rasmussen2000infinite,neal2000markov,medvedovic2002bayesian,medvedovic2004bayesian,mcdowell2018clustering} with the difference that we have placed here multiple independent mixtures over the prior distribution of the fPC scores. In practice we used the truncated stick-breaking construction and tested the model with different commonly chosen values of  $ J $ ($ J=20,30 $ and $ 50 $). The upper bound $ J $ should be chosen sufficiently large to ensure $ J_{k}^{+}< J $ in each eigendimension. Larger $ J$s will naturally impact on computations (e.g. in our applications we observed the computational time of the model with $ J=50 $ to be $ \sim1.5 $ higher than with $ J=20 $).
All the conditional posteriors of this model (most of them available in closed form) are provided in Appendix A. Markov chain Monte Carlo (MCMC) techniques are used to simulate from the joint posterior distribution of all parameters given the data. Reconstruction of the smooth trajectories $ x_{it} $ is made easy by its linear relationship with the model parameters $ \xi_{ik} $; thus it is possible to obtain the posterior distribution of the $ i $-th curve for every $ t $ and at every MCMC iteration $ w $,
\begin{equation}\label{eq9}
x_{it}^{(w)}= \bar{x}_{t}+\sum_{k=1}^{K}\xi_{ik}^{(w)}\phi_{kt},\,\,\,i=1,\dots,n;\,\,t=1,\dots,T,\,\,w=1,\dots,W,
\end{equation}
where $ \bar{x}_{t} $ is the smoothed estimate of the sample mean $\sum_{i=1}^{n}y_{it}/n $. It follows that symmetric $ 95\% $ point-wise credible intervals for each trajectory-specific mean can be obtained easily from Equation~(\ref{eq9}) by considering the $ (1-\alpha)/2 $ and $ \alpha/2 $ quantiles of the $ \left\lbrace x_{it}^{(1)},\dots,x_{it}^{(W)} \right\rbrace  $ empirical distribution.

\subsection{Clustering}
\label{sec2.3}
In this section we focus on the clustering of fPC scores. The discrete nature of the DP is very useful for clustering as it allows ties among the latent $ c_{ik} $ \cite{muller2015bayesian}; therefore, DP mixtures implicitly return classification through the allocation of each fPC score to a generating distribution with some probability.
Clustering uncertainty can be evaluated at different levels such as the number of clusters, the size of each cluster and the fPC scores assigned to them. For the explorative purpose of our model we avoid the use of automated algorithms to select a final partition of the fPC scores (either classical hierarchical or partitioning algorithms based on the similarity matrix \cite{medvedovic2004bayesian} or more recently proposed algorithms based on a loss function over clusterings \cite{wade2018bayesian}). Instead, we propose a 3-step exploration of the empirical distribution of generated clusterings which we find useful to evaluate clusters uncertainty arising from the data. After burn-in, the empirical distribution of generated clusterings $\left\lbrace  \boldsymbol{c}_{k}^{(1)},\dots,\boldsymbol{c}_{k}^{(W)}\right\rbrace$ can be considered a good approximation of the true posterior distribution \cite{neal2000markov} and it can be used to obtain other distributions of interest, such as the number and size of non-empty clusters, maximum a posteriori probabilities (MAPs) and pairwise probability matrices (PPMs). We make use of these distributions in a 3-step exploration.

\textit{Step 1:} The distribution of the number of non-empty clusters $ J_{k}^{+} $ can be obtained by exploring the values of the classification variable $ \boldsymbol{c}_{k} $ for all the $ W $ iterations retained after burn-in $\left(   J_{k}^{+,w}=\max_{j}\left\lbrace \boldsymbol{c}^{w}_{k}\right\rbrace \right) $. Although considering the number of non-empty clusters $J_{k}^{+}$ does not account for size and stability (i.e. the number of times a cluster appears in the MCMC chain), the distribution of $J_{k}^{+}$ provides a useful first check for assessing the presence of more than one cluster in each eigendimension. For this purpose, we used the Bayes Factor (BF) defined as $ \left\lbrace P_{\pi}(J^{+}_{k}=1)/P_{\pi}(J^{+}_{k}>1) \right\rbrace \times \left\lbrace P(J^{+}_{k}>1)/ P(J^{+}_{k}=1)  \right\rbrace   $ where $ P_{\pi}(J^{+}_{k}=j) $ denote posterior probabilities and $ P(J^{+}_{k}=j)$ the relative prior probabilities which can be obtained by simulating from the prior distribution of $ \boldsymbol{c}_{k} $.
A BF greater than 1 suggests absence of clusters in the fPC scores of a specific eigendimension; hence, this step identifies those eigendimensions where clusters are more likely to exist in the data.

\textit{Step 2:} The distribution of the cluster size can be obtained by counting for each iteration $ w $ the number of fPC scores allocated to the same label $ \left( \sum_{i=1}^{n} \text{I}\left( c^{(w)}_{ik}=j\right) ,\,\forall j\in\left[ 1,J^{+}_{k}\right] \right) $ or by monitoring the posterior distribution of the mixing proportions $ p_{jk} $. Although there is no guarantee that fPC scores joining a cluster remain loyal to it, the size of clusters permits the identification of clusters which are populated only sporadically as a result of the uncertainty in the classification of subsets of fPC scores. The distribution of these clusters has typically a notable probability mass at zero. Therefore, this second step can help understand the number and dimension of clusters we expect to see in each eigendimension and the relative uncertainty. 

\textit{Step 3:} Finally, MAPs and PPMs can help refine our understanding of the underlying clustering. MAPs are commonly used to identify the most probable clustering for each observation and they can be computed by identifying for each fPC score the posterior mode of $ c_{ik} $ from the empirical distribution of generated clusterings. MAPs are known to be limited by the possible presence of multiple modes and cases where individuals who share the same modal group are less frequently together than with others in different clusters. These issues can be addressed by the PPMs which represent the posterior belief for all pairs of curves to belong to the same cluster regardless of the clustering label.\cite{medvedovic2002bayesian,medvedovic2004bayesian,mcdowell2018clustering} For each iteration $ w $, an $ n\times n $ association matrix $ \delta(\boldsymbol{c_{k}}) $ can be obtained with indicators $ \delta_{ii^{'}}(\boldsymbol{c_{k}}) $ which takes value 1 if fPC score $ i $ and $ i^{'} $ in eigendimension $ k $ are clustered together and 0 otherwise. Element-wise averaging over all these association matrices yields the PPM. Combining the exploration of MAP and pairwise probabilities can narrow down a decision on the most likely partition of the fPC scores. 

Although we find limitations for each of these steps individually to draw robust conclusions, considering them together as a whole provides rich information on the (a posteriori) most likely partition for each eigendimension. Particularly in the case of complex phenomena, such as those captured by neuroscientific recordings, a thorough exploration of cluster uncertainty in the data should be always considered to ensure a sensible interpretation of the results.
We present an application of these analyses to fMRI and EEG data in Section \ref{sec4}. In a Bayesian mixture model where cluster identification is of interest, extra care should be taken to avoid label switching arising from the symmetry in the likelihood of model parameters. This can be avoided either by imposing identifiability constraints on the parameter space or by employing relabelling algorithms. In our simulation study and applications we found that imposing constraints on the order of cluster means ($ \mu_{1k}<,\dots,<\mu_{Jk} $) or weights ($ p_{1k}<,\dots,<p_{Jk} $) was enough to successfully control label switching. 

\subsection{fPC score clustering as generalisation of standard clustering}
\label{sec2.4}
In the standard infinite mixture model based clustering, the indicators $c_{i}=c_{i'}=j$ with $i\ne i'$ would associate a couple of trajectories to a certain cluster $ j$ with probability $P_{ii'}$. On the other hand, by placing infinite mixtures over the fPC scores for every eigendimension retained, we allow for a more complex network of dependence among curves. In our model, $c_{ik}$ and $ c_{i'k}$ would associate fPC scores $i$ and $i'$ to potentially different clusters in every eigendimension $k$ with probability $P_{ii'k}$. It follows that a pair of curves could happen to share the same cluster in only part of the $ K $ eigendimension retained, expanding the standard model based clustering to a richer classification method. Furthermore, as each dimension represents a mode of variation (eigenfunction) and its importance (eigenvalue), our method offers additional insights into the underlying spatio-temporal structure of the data.
In the following sections we show how clustering fPC scores produces a rich spatio-temporal exploration of complex neuroscientific data.

\section{Simulation study}
\label{sec3}
\subsection{Simulation scenarios}
\label{sec3.1}

We performed a simulation study to assess the performance of PCl-fPCA model and compare it to the standard Bayesian fPCA model in terms of both curve reconstruction and classification for different data generating processes and noise levels. We also included for comparison two frequentist approaches: the standard fPCA model \cite{ramsay2005functional} and a modified version of the model by Liu et al. \cite{liu2017functional} that we adapted to the features of neuroscientific data. In this latter model, curve dependence is captured through the fPC scores by means of independent Mat\'{e}rn functions for each eigendimension retained.

\begin{figure}
	\centering
	\includegraphics[width=0.85\linewidth]{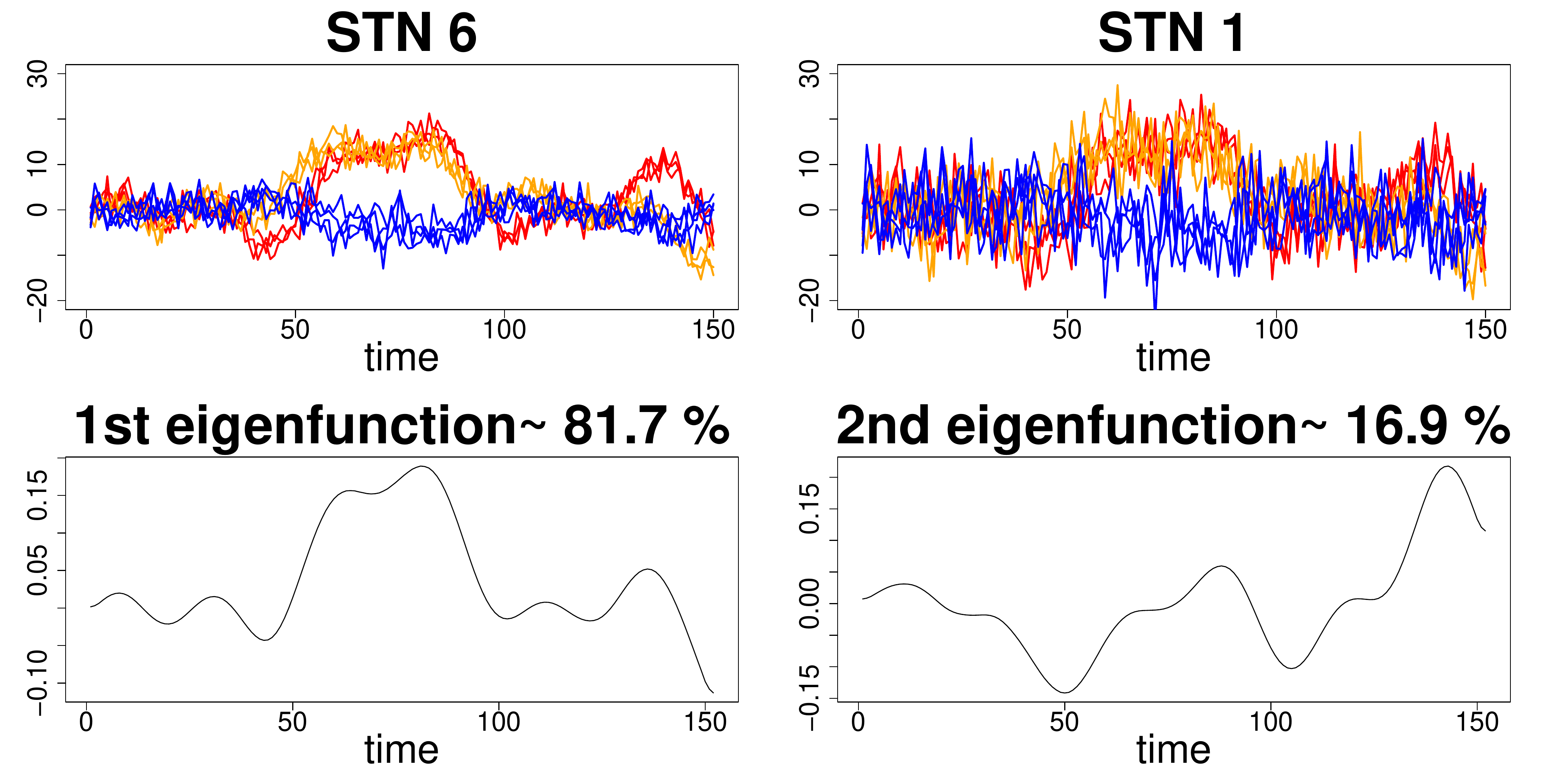}
	\caption{Simulation study: (Top) an example of curves from DGP1 with low random noise (STN6) and high random noise (STN1). (Bottom) the first and second eigenfunctions extracted from a set of DGP1 curves with STN6. This figure appears in colour in the electronic version of this article.}
	\label{fig:1} 
\end{figure}

In order to test model performance with simulated data matching those of the targeted neuroscientific applications as closely as possible, we generated two eigenfunctions from simulated data resembling evoked responses in the brain using the function \texttt{pca.fd} from the \texttt{fda} package in \texttt{R} \cite{fdapackage}. Subsequently, we defined three data generating processes (DGP) that differ in the way the fPC scores are generated: in the first DGP (DGP1), scores are generated from different mixtures of Gaussian distributions in the two eigendimensions considered; in the second DGP (DGP2), fPC scores dependence in the first eigendimension is generated from a Mat\'{e}rn function while in the third DGP (DGP3), dependence of fPC scores is generated by independent Mat\'{e}rn covariance functions with different parameter values in each eigendimension. For each DGP, we combined the two eigenfunctions with the fPC scores to build the simulated datasets. We applied a random Gaussian noise and tested the models with both high and low signal-to-noise ratios (STN=6 and 1 respectively). Figure~\ref{fig:1} shows an example from the set of $ 100 $ generated curves in DGP1 where either a low or high random noise is added.

One hundred datasets ($ L=100 $) for each DGP and STN were input to fPCA first for curve smoothing using cubic B-splines and dimension reduction by estimating the respective eigenvalues and eigenfunctions using the function \texttt{pca.fd} from the \texttt{fda} package in \texttt{R} \cite{fdapackage}. We retained a number of dimensions $ K $ explaining at least $ 95\% $ of the total variability in curves. Figure~\ref{fig:1} shows eigenfunctions and their weights extracted after smoothing a set of low-noise curves for the first DGP.

We adapted the general model presented in Section \ref{sec2.2} to the specific simulation analysis using eigenvalues $ \lambda_{k} $ and their properties to develop vaguely informative prior distributions for the parameters $ r,\,\beta $ and $ Q $ (Equations (\ref{eq7}) and (\ref{eq8})) in the two eigendimensions retained $ k=1,2 $.
We set $ r\in\left\lbrace 1/\hat{\lambda}_{1},\,1/\hat{\lambda}_{2}\right\rbrace  $ and $ Q\in\left\lbrace 10,5\right\rbrace  $ as well as setting $s_{j,1}\sim\Gamma(1,\lambda_{1}) $ and $\sigma_{j2}\sim \text{U}[0,\sqrt{\lambda_{2}}] $. The use of a uniform distribution in the second dimension favours the search of smaller clusters than in the first eigendimension, as increasingly local features should be expected in trailing modes of variation.\cite{liu2017functional} We made sure that even the smallest upper-bound $ Q $ of the dispersion parameter $ \alpha $ distribution represented an expected number of clusters a priori far higher than the ground truth.\cite{escobar1994estimating,jara2007dirichlet} A similar choice for $ \alpha $ was specified by De Iorio et al.\cite{de2018bayesian} due to the resulting stable computations.

We coded the model in \texttt{R} using the \texttt{rjags} package\cite{rjagsplum}, and employed a conservative approach using $ 100,000 $ iterations for the burn-in and retaining the subsequent $ 100,000 $ MCMC iterations.\cite{medvedovic2002bayesian,gelman2013bayesian} The convergence diagnostics did not suggest lack of convergence for all the parameters of interest. We used a thinning of 5 to store results from 100 simulated datasets efficiently (approximately 70 MB each with $ K=2 $). It takes 36 minutes on average to complete one simulation run on a 2-core Intel CPU running at 2.7 GHz with 8 GB RAM.

We used Integrated Mean Squared Error (IMSE) to measure and compare reconstruction performance between PCl-fPCA model and the competitor models. IMSE and its associated approximation for every curve $ i $ are given by
\begin{align}\label{eq13}
\text{IMSE}_{i}=\text{E}\bigg\{\int\big(\widehat{x}_{it}-x_{it}\big)^2 \text{dt}\bigg\}\approx\frac{1}{L}\sum_{l=1}^{L}\bigg\{\frac{1}{T}\sum_{t=1}^{T}\big(\widehat{x}_{i_{l}t}-x_{it}\big)^{2}\bigg\},
\end{align}
where the expectation is taken with respect to the underlying curve $ x_{i} $. The IMSE is a useful measure of performance in density estimation and is frequently used in curve reconstruction. \cite{gentle2009computational,rasheed016bayesian} In addition, as curves correlation $ \rho_{ii'} $ is often of interest in neuroscientific applications (e.g. for measuring the degree of functional connectivity between brain areas), we measured correlations reconstruction using the L2 norm $ ||\widehat{\rho}_{ii'}-\rho_{ii'} ||_{2} $ and compared it with those of the competitor models.

In order to assess the proposed model clustering performance in DGP1, we adopted the Adjusted Rand Index (ARI) to quantify the similarity between the estimated partitions (using MAP) and the ground truth for every simulated dataset $ l $ and eigendimension $ k $. The ARI is commonly used in the literature to assess clustering performance as it varies between exact partition agreement (1) and when partitions agree no more than is expected by chance (0).\cite{hubert1985comparing, mcdowell2018clustering} Moreover, we measured the improvement in distance ($L_{2}$ norm) between the posterior pair-wise probability matrices and the ground truth to evaluate the clustering performance of PCl-fPCA model by taking into account cluster uncertainty.
Further details on the simulations setting can be found in WebC section of the Supplementary Material.

\subsection{Simulation results}
\label{sec3.2}
\begin{figure}
	\centering
	\includegraphics[trim={2cm 0 2cm 0},width=1.0\linewidth]{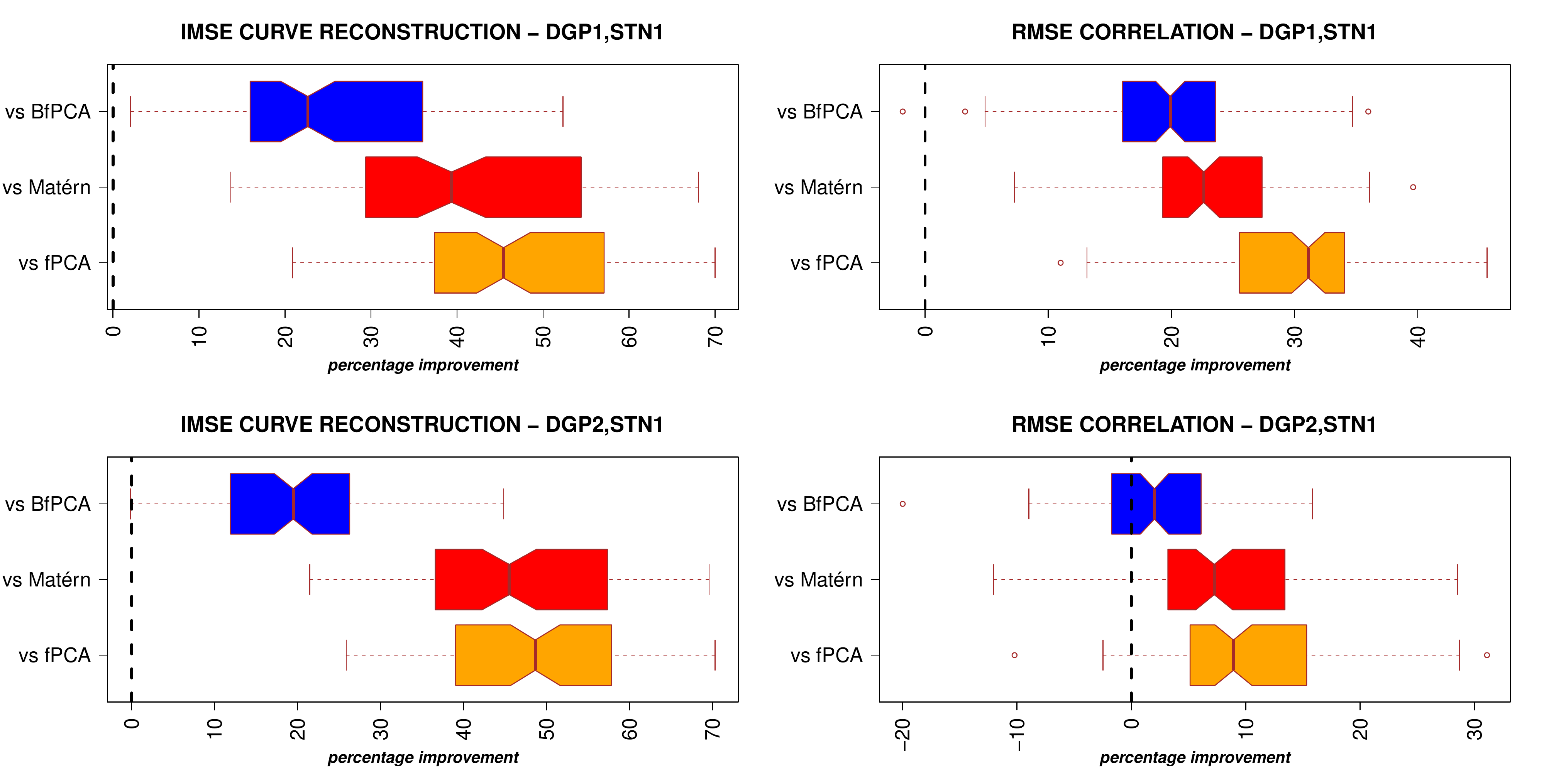}
	\caption{Simulation study: curve and correlation reconstruction for Data Generating Processes (DGP) 1 and 2 with high noise (STN1). IMSE and RMSE improvement percentage using PCl-fPCA model versus standard Bayesian fPCA (BfPCA), fPCA model for correlated curves (Mat\'{e}rn) and standard fPCA model (fPCA). This figure appears in colour in the electronic version of this article.\label{fig:2}}
\end{figure}

Results of curve and correlation reconstruction are reported in Figure~\ref{fig:2}.
The case where STN = 1 is particularly relevant because neuroscientific data are usually affected by high noise. In this scenario, PCl-fPCA model highly improved curve reconstruction compared to all competitor models as $ 100\% $ of the true curves were better recovered under PCl-fPCA and the median improvement in IMSE ranged from $ 22\% $ to $ 45\% $. Moreover, a similar improvement was also obtained for DGP2 where clustering is present in only one eigendimension (Figure~\ref{fig:2}, bottom left). In addition, correlation reconstruction was also better achieved under PCl-fPCA with a median percentage of improvement ranging from $ 20\% $ to $ 30\% $ for DGP1 and $ 2\% $ to $ 8\% $ for DGP2 (Figure~\ref{fig:2}, right column). In the case of low noise (STN6), the proposed model still performed better than the competitors for DGP1 and achieved values of IMSE and RMSE similar to those of the best competitor models in DGP2 (Supplementary Material, WebB, Figure A2).
Interestingly, even when no clusters are expected in both eigendimensions (DGP3), the performance of the PCl-fPCA was still comparable to the best ones achieved by competitor models for both low and high noise levels (WebB, Figure A3).

The performance of the PCl-fPCA model in terms of classification is reported in Table~\ref{Tab3}.
\begin{table}
	\centering
	\caption{Simulation study: clustering performance of PCl-fPCA in DGP1. The table reports median and interquartile range of ARI computed for each simulated dataset and every STN and eigendimension analysed.\label{Tab3}}
	{\tabcolsep=5.25pt	\begin{tabular} {@{}lc@{}}
			\hline Eigendimension & $ARI $\\
			\hline
			\vspace{4.5pt}		
			\textbf{STN=1}    & \\ 	
			\textit{1st dim} & 1 [1,1]  \\ 
			\vspace{2.5pt}
			\textit{2nd dim} & 0.753 [0.444,0.868] \\ 
			\hline
			\vspace{4.5pt}	
			\textbf{STN=6}    &  \\ 	
			\textit{1st dim} & 1 [1,1] \\
			\vspace{2.5pt}
			\textit{2nd dim} & 0.966 [0.933,0.966] \\ 
			\hline\hline
	\end{tabular}}
\end{table}
The proposed model scored high in the ARI classification index in both eigendimensions studied; two and three clusters were expected in the first and second dimension respectively in DGP1. Clusters in the first eigendimension were always correctly identified by ARI for both high and low signal to noise ratios. The identification of three clusters in the second eigendimension was more challenging as they were smaller and nearer to each other; however, scores near 1 were almost always obtained when the low noise scenario was tested and even in the case of high noise we observed fairly high scores. Similar results were achieved by measuring the improvement in distance (L2 norm) between the posterior pair-wise probability matrices and the ground truth  to account for cluster uncertainty in the classification performance (WebC, Table A3).

Figure A4 in section WebB of the Supplementary Material provides evidence of the improved level of information achieved by PCl-fPCA in the DGP1 scenario. Overall, PCl-fPCA model outperformed the competitors in curve reconstruction under different data generating processes, especially in the case of high noise in the data; moreover, for the case where clusters are not limited to one eigendimension, the proposed model was able to retrieve the original spatial partition in each eigendimension and bring to light important relationships between clusters. These results could further help the understanding of underlying neuroscientific phenomena in a real data scenario.

\section{Application}
\label{sec4}
In this section we present two applications of the PCl-fPCA model to the analysis of neuroscientific data from fMRI and EEG recordings. In Sections~\ref{sec4.1} and \ref{sec4.2}, the PCl-fPCA model is used to explore underlying brain patterns arising from a short time window fMRI recording of a healthy subject at rest. In the emerging field of dynamic functional connectivity, the analysis of the evolution of brain patterns within a short time window is of particular interest as it could uncover transient configurations of coordinated brain activity.\cite{chen2017methods} The aim of the present fMRI analysis is to verify whether the results obtained on a short time window recording (1 minute) are in line with the current knowledge on brain resting-state networks obtained from static functional connectivity studies where results are typically averaged over 5-15 minutes recordings. In Sections~\ref{sec4.3} and \ref{sec4.4} the PCl-fPCA model is used for artefacts identification in the EEG recording of a healthy subject under a two-stimuli paradigm (match vs unmatch images). The presence of artefacts originating from sources different from the brain and contaminating brain signals is a well-known problem in EEG recordings and an active area of research in neurophysiology. \cite{islam2016methods} The aim of the present EEG analysis is to check whether the fPC-PCA model can be successfully used to identify the spatio-temporal features of different artefacts and the location of the relative affected brain areas.
\subsection{fMRI setting}
\label{sec4.1}

The study relates to a thirty-year-old healthy woman volunteer who underwent a resting-state fMRI at the Department of Radiology, Scientific Institute Santa Maria Nascente, Don Gnocchi Foundation (Milan, Italy) during February 2015. 
The recording was carried out  using a 1.5 T Siemens Magnetom Avanto (Erlangen, Germany) MRI scanner with 8-channel head coil. The subject was asked to lie down in the MRI machine in supine position with eyes closed while Blood Oxygenation Level Dependent Echo Planar Imaging (BOLD EPI) images were acquired. She was instructed to keep alert and relaxed; no specific mental task was requested.

High resolution T1-weighted 3D scans were also collected to be employed as anatomical references for fMRI data analysis. Standard pre-processing involved the following steps: motion and EPI distortion corrections, non-brain tissues removal, high-pass temporal filtering (cut-off $ 0.01 $ Hz) and artefacts removal using the FMRIB ICA-based Xnoiseifier (FIX) toolbox. \cite{griffanti2014ica}
After the pre-processing, the resulting 4D dataset was aligned to the subject's high-resolution T1-weighted image, registered to MNI152 standard space and  resampled to $ 2\times 2\times 2\,\, mm^{3}$ resolution. One minute length series (sampled at 0.5 Hz) were extracted as the average signal within each of 90 regions of interest (ROIs) according to the Automated Anatomical Labeling (AAL90) coordinates.
The resulting $ 30\times90 $ dataset was input to fPCA for curve smoothing and dimension reduction using the \texttt{pca.fd} function from the \texttt{fda} package in \texttt{R}.\cite{fdapackage} The set of 90 smooth curves and the retained eigendimensions are shown in Figure A5 of Supplementary Material WebB. We kept the first three dimensions explaining more than $ 85\% $ of the total variability while accounting for more than $ 10\% $ each.

We adapted the general model in Section~\ref{sec2.2} following the approach taken in the simulation study (Section~\ref{sec3.1}), favouring global patterns in the first eigendimension and local patterns in the remaining dimensions. We assessed convergence using trace plots and BGR diagnostics and the number of independent retained samples by computing the effective sample size (WebD, Supplementary Material). We employed the same computational approach described in Section~\ref{sec3.1} and it took 59 minutes to run the analysis with $ K=3 $ on a 2-core Intel CPU running at 2.7 GHz with 8 GB RAM. Furthermore, we carried out a sensitivity analysis by varying the values of the hyperparameters $\beta, Q $ and the distribution of $ s $  in each dimension (WebA, Supplementary Material).

\subsection{fMRI analysis results}
\label{sec4.2}
The posterior probabilities associated with the single cluster (i.e. no clusters) scenario were $ 0.012,\,0.124 $ and $ 0.058 $ for the three eigendimensions $ k $, respectively. The Bayes factors for the first eigendimension was $ 0.53 $, which indicates some evidence against no clusters. Conversely, the second and third dimensions returned $\text{BF}=2.93 $ and $ 1.33 $ respectively, which can be interpreted as evidence in favour of a single cluster.  It is worth noting that, as the implied prior probabilities were highly in support of multiple clusters, the BF for $ k=2 $ and $ 3 $ show a diametrical change from prior to posterior belief. These results are also confirmed by a BF sensitivity analysis which is reported in the supplementary material (WebA).

Figure~\ref{fig:3} shows the posterior probability for a cluster being empty and the posterior distributions of cluster size given it is not empty.
\begin{figure}
	\centering
	\includegraphics[trim={5cm 0 5cm 0},width=0.8\linewidth]{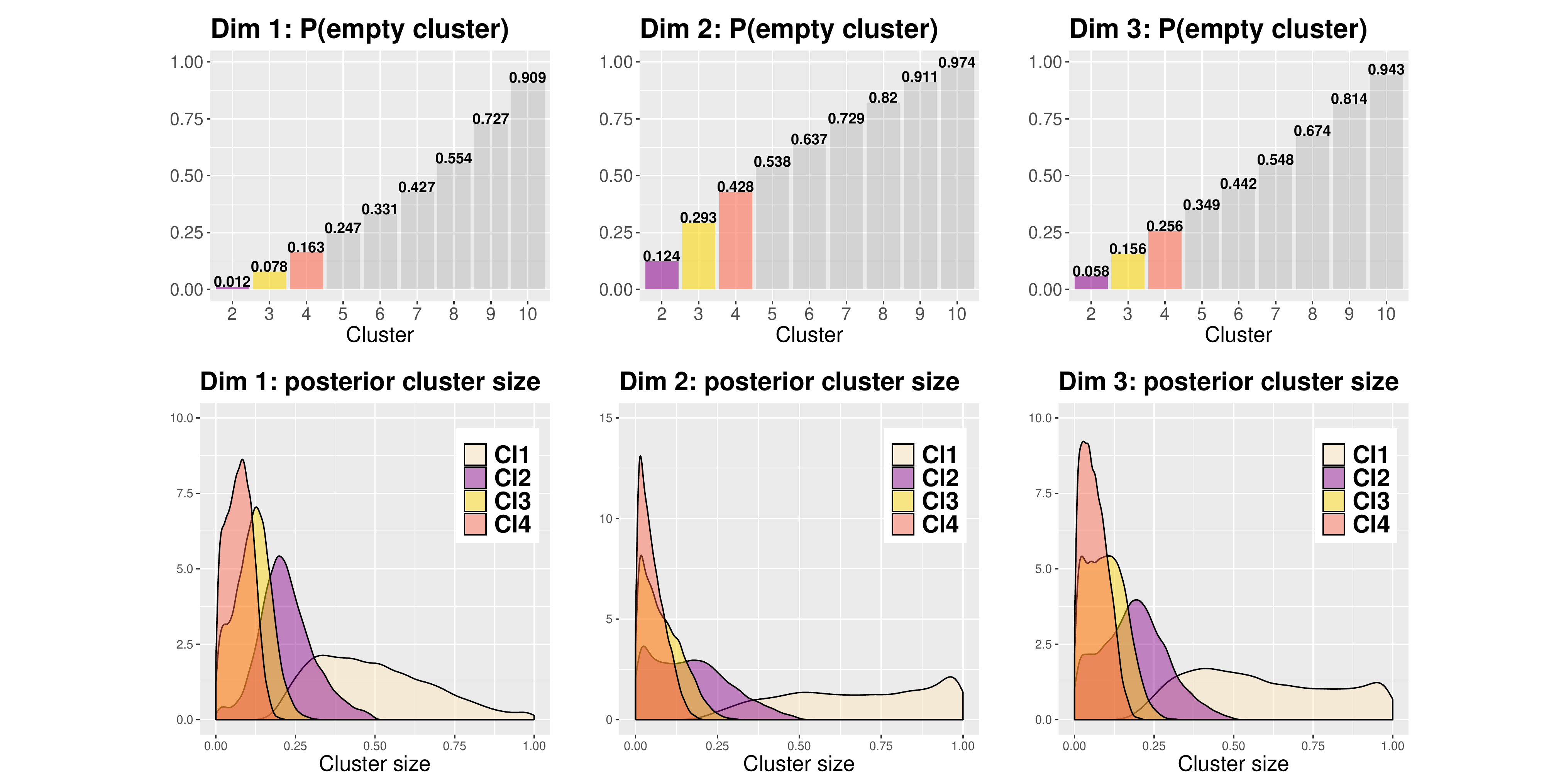}
	\caption{fMRI data analysis: cluster identification. The first row shows the posterior probabilities of being empty for the second to tenth clusters in the three eigendimensions (Dim 1:3) analysed. The second row shows the posterior distributions of cluster size (given it is not empty) among the first four clusters (Cl1:Cl4, right to left). This figure appears in colour in the electronic version of this article. }
	\label{fig:3} 
\end{figure}
Two to three clusters seem to emerge in dimension 1; the size of the second cluster (Cl2, second from the right in Figure~\ref{fig:3}, bottom-left panel) has a peak around $ 20\% $, very small mass near zero, and a very low probability of being empty. The third cluster (Cl3) has a size peaking at $ 12\% $ but more mass near zero and a  higher probability of being empty. On the other hand, dimension 2 and 3 seem to suggest the presence of no more than one cluster each. The second cluster in both these dimensions has higher probability of being empty and the distributions of size have much more mass around zero. Furthermore, the distributions of the first cluster (Cl1) in both dimensions have a notable peak around $ 90\% $ suggesting that, even when more than one cluster is considered, the large majority of fPC scores in dimension 2 and 3 tends to be gathered within a single large cluster.

The use of MAPs suggests there might be no more than 2 groups in the first dimension and 1 group in the second and third dimensions. Clustering with MAPs in the first dimension identified $ 9\% $ of curves whose trajectories are wigglier and with a visibly shorter inter-peak difference between the first positive and negative peaks compared to the other group (WebB, Figure A6). Figure A7 of section WebB in the Supplementary Material shows an example of curve reconstruction using the posterior mean and $ 95\% $ point-wise credible bands of the subject specific mean. Curves in cluster 2 pertain to brain areas from the occipital lobe (Calcarine, Cuneus, Lingual, Inferior Occipital Gyrus) and parietal lobe (Precuneus).

\begin{figure}
	\centering
	\includegraphics[trim={10cm 0 7cm 0},width=0.74\linewidth]{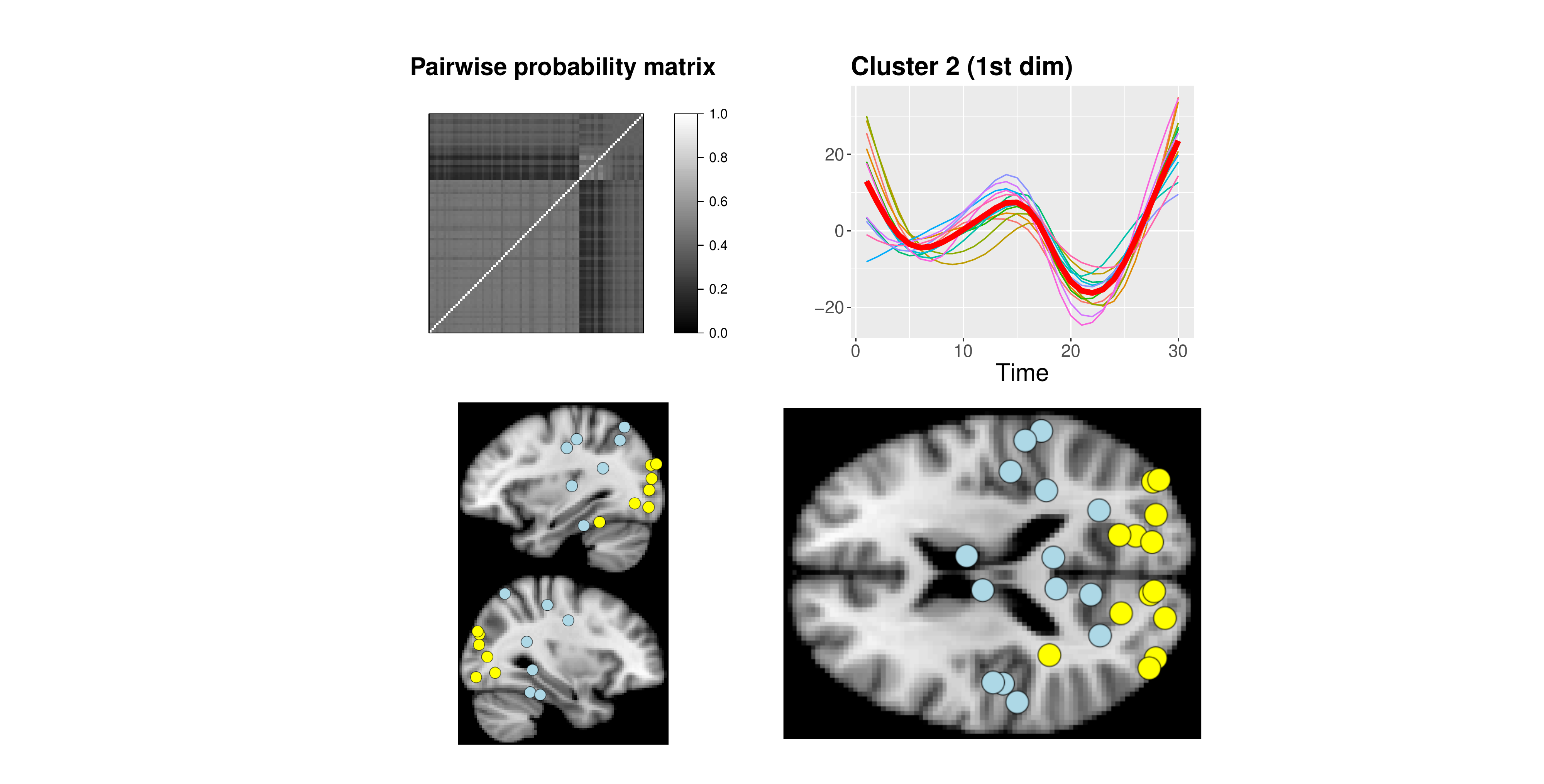}
	\caption{fMRI data analysis: cluster identification with pairwise probabilities. Top-left: pairwise probabilities suggesting a tripartition of curves in the first eigendimension. Top-right: cluster 2 updated according to the partition suggested by pairwise probabilities. The thick line represents the cluster mean. Bottom: the 3-D representation of clusters 2 and 3 over sagittal and axial slices of the human brain, where yellow (light) dots represent locations in cluster 2 and blue (dark) dots those in cluster 3. This figure appears in colour in the electronic version of this article.}
	\label{fig:4} 
\end{figure}

By analysing the pairwise probability matrix, a more comprehensive classification emerged. The previously dichotomous partition in dimension $ k=1 $ is now enriched by a third group of brain areas with no clear clustering preference (grey band at the top-right of the pariwise probability matrix in Figure~\ref{fig:4}).
Cluster 2 comprises $ 16\% $ of curves which all represent areas from the occipital lobe (yellow-light dots), while curves in cluster 3 (blue-dark dots) belong to the cingulate cortex (Middle and Posterior Cingulate Cortex), parietal (Parietal Superior Lobule, Precuneus) and temporal (Middle and Inferior Temporal Gyrus) lobes (Figure~\ref{fig:4}, a colour version of this figure can be found in the online version of the article).

We note that these three clusters are supported in the neuroimaging literature. It is well established that primary and extra-striate visual regions are active at rest \cite{van2010exploring} and have a role in processing mental imagery.\cite{zhang2018intrinsic} Just outside the visual cortex, the Temporal Inferior Gyrus takes part to the visual ventral stream which links information from the visual cortex to memory and recognition.\cite{milner2017two}
Moreover, the Posterior Cingulate Cortex is known to interact with several different brain networks simultaneously and it participates in the Default Mode Network together with part of the parietal lobe.\cite{leech2012echoes} Conversely, it has been suggested that areas pertain to the Prefrontal Cortex (all included in cluster 1) have less long-range connectivity in the resting state condition.\cite{tomasi2011functional}
Finally, the sensitivity analysis further confirmed our findings as they were robust to changes in both shape and value of the hyperparameters (WebA, Supplementary Material).

\subsection{EEG setting}\label{sec4.3}
	For our second application we employed data from an EEG study on brain activations following object recognition tasks (Event Related Potentials, ERPs).\cite{zhang1995event} ERPs are very small bio-electrical signals generated by the brain in response to specific events or stimuli. They are EEG changes time locked to motor, sensory or cognitive events that provide a non-invasive approach to study psychophysiological correlates of mental processes.\cite{sur2009event} In contrast, body or eye movements introduce large artefacts to EEG recordings and trials contaminated with artefacts need to be corrected or even discarded.\cite{plochl2012combining} In the present study we employed the PC-fPCA model for artefacts identification in the EEG recording of a single healthy subject. The individual was presented with two separate stimuli in the forms of images taken from the 1980 Snodgrass and Vanderwart picture set.\cite{snodgrass1980standardized} The second stimulus was either a different image (unmatch) or the same image (match) as in the first stimulus. We used the data-driven clustering of the PCl-fPCA model to identify the spatio-temporal features of different artefacts and the relative affected brain areas. 
	  
The data were recorded using a cap with 64 electrodes  placed on the subject's scalp and the brain activity at each recording electrode was sampled at 256 Hz for 1 second. Further details on the recording setting can be found in Zhang et al.\cite{zhang1995event} We
considered both the unmatched and matched tasks within the same analysis and used our PCl-fPCA model to find data-driven differences in the morphology of the curves. Therefore, a  $ 128 \times 256 $ dataset was input to fPCA for curve smoothing and dimension reduction using the \texttt{pca.fd} function from the \texttt{fda} package in \texttt{R}.\cite{fdapackage} The set of 128 smooth curves and the retained eigendimensions are shown in Figure A8 of Supplementary Material WebB. We kept the first two dimensions explaining  $ 90\% $ of the total variability while accounting for more than $ 10\% $ each. We applied the same model settings described in Section~\ref{sec4.1}; we assessed convergence using trace plots and BGR diagnostics and the number of independent retained samples by computing the effective sample size (WebD, supplementary material). We employed the same computational approach described in Section~\ref{sec3.1} and it took 64 minutes to run the analysis with $ K=2 $ on a 2-core Intel CPU running at 2.7 GHz with 8 GB RAM.

\subsection{EEG analysis results}\label{sec4.4}	
\begin{figure}
	\centering
	\includegraphics[width=0.6\linewidth,trim={9.8cm 0 9.0cm 0.35cm},clip]{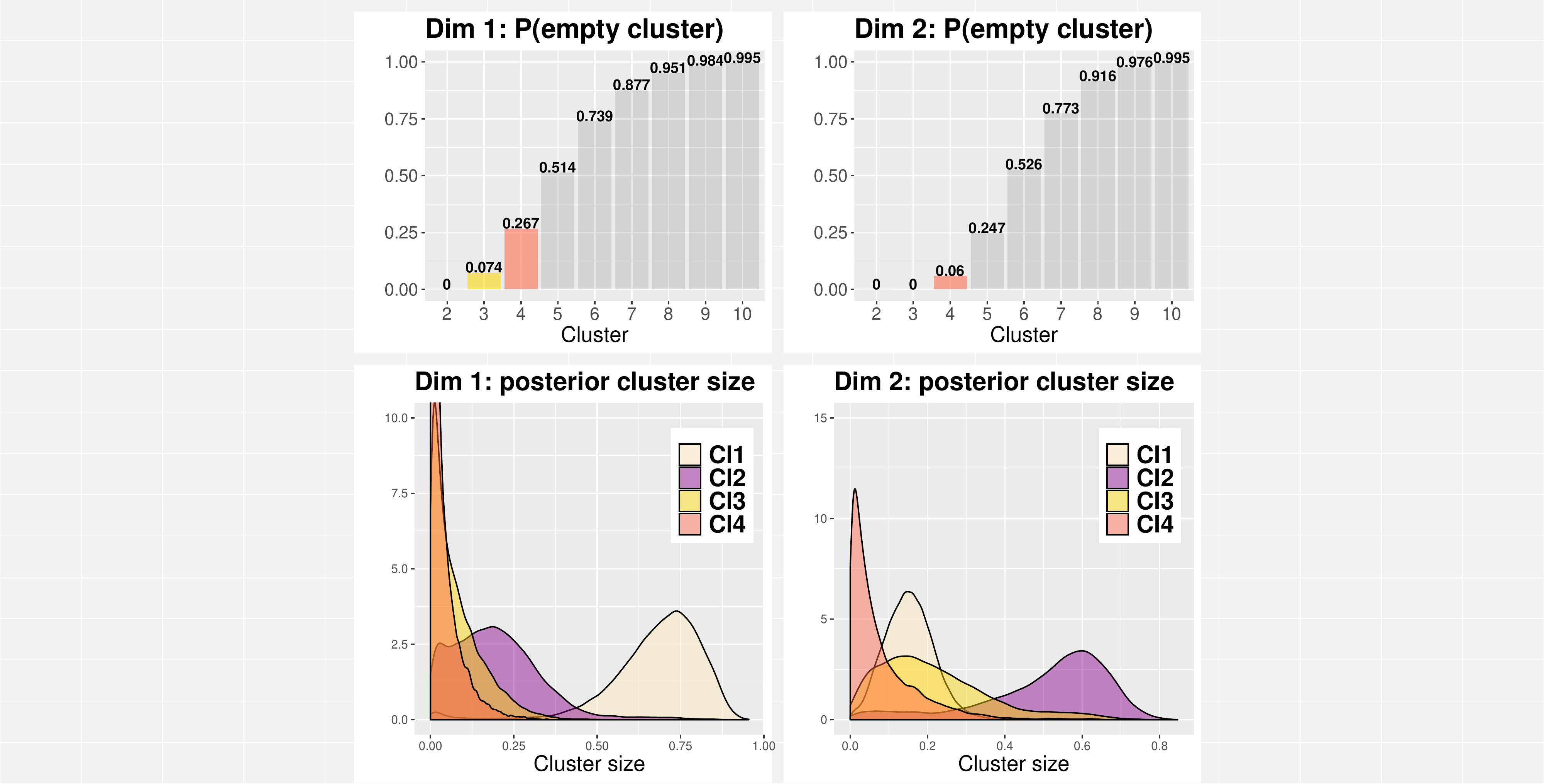}
	\caption{EEG data analysis: cluster identification. The first row shows the posterior probabilities of being empty for the second to tenth clusters in the two eigendimensions analysed. The second row shows the posterior distributions of cluster size (when not empty) among the first four clusters (Cl1:Cl4, right to left). This figure appears in colour in the electronic version of this article. }
	\label{fig:5} 
\end{figure}

Two clusters seem to emerge in dimension 1. The size of the second cluster (Cl2, second from the right in Figure~\ref{fig:5}, bottom-left panel) has a peak around $ 20\% $, and a low probability of being empty. The third and fourth clusters (Cl3, Cl4) have both sizes peaking near zero and higher probabilities of being empty. On the other hand, dimension 2 clearly indicates the presence of three clusters with sizes peaking at $ 60\% $, $ 20\% $ and $ 20\% $ and very low probabilites of being empty. Furthermore, the distributions of the first cluster (Cl1) in both dimensions have very low mass near 1, supporting the presence of multiple clusters in both dimensions.
 	
\begin{figure}
	\centering
	\includegraphics[width=0.85\linewidth,trim={6cm 4.1cm 4.8cm 3.1cm},clip]{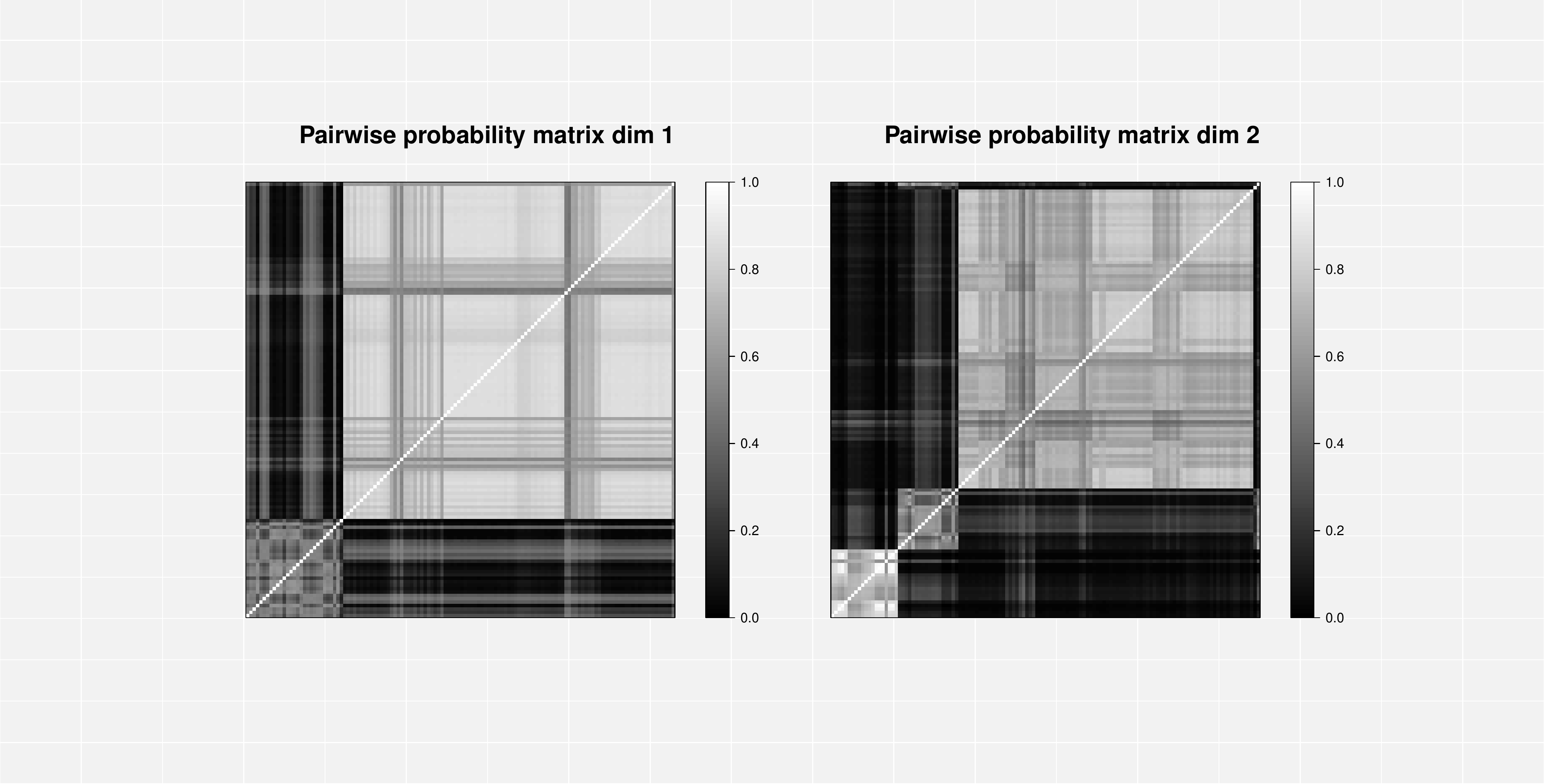}
	\caption{EEG data analysis: cluster identification with pairwise probabilities. First column: pairwise probabilities suggesting a bipartition of curves in the first eigendimension (top) and a tripartion in the second (bottom).}
	\label{fig:6} 
\end{figure}

Both MAP and pairwise probability analyses confirmed the presence of 2 clusters in the first dimension and 3 clusters in the second dimension (Figure~\ref{fig:6}). The second cluster in the first eigendimension contains all the recordings from electrodes in the frontal areas for both the matched and unmatched tasks (Figure~A9, WebB, Supplementary Material). These curves have a marked peak at the end of the recording, indicating a possible artefact (probably originated from eye blinking), and they appear to have two separate underlying patterns. These trends are captured in the clustering of the second eigendimension where the second and third clusters further divided the EEG activity in the frontal brain areas between those recorded during the matched and unmatched tasks (Figure~A9, WebB, Supplementary Material). Notably, despite all curves showing more variability toward the end of the recordings, we found that only those from frontal areas have a consistently different behaviour from that of the group. This is in line with the work of Zhang et al.\cite{zhang1995event}, where the authors excluded frontal region recordings from part of their analyses because of an inconsistent wave morphology compared with the wave form of the other regions. Frontal areas are known to be prone to recording artefacts particularly from eye movement which might have affected the different wave forms observed in these data. \cite{plochl2012combining} Furthermore, the data-driven separation of frontal area curves into tasks (matched and unmatched) suggests the effect of two separate artefacts on the amplitude of these recordings.

\section{Discussion}
\label{sec5}

The processing of the human brain is a complex phenomenon in both time and space. The modelling of spatio-temporal datasets in the big data era is a challenge becoming every day more demanding as we struggle to keep up with the overwhelmingly larger datasets we are required to make sense of.
Moreover, the extraordinary advancements in neuroimaging of the last decades have focused large part of neuroscientists and statisticians' efforts on the spatial domain both in clinical practice and research (see, for example, Durante et al.\cite{durante2018bayesian}). Nevertheless, the study of how interactions among brain regions change dynamically during an experiment, (i.e. \textit{dynamic functional connectivity}) has recently attracted
interest in the neuroimaging literature.\cite{warnick2018bayesian}  
In fact, the time domain retains important neurophysiological information on brain functioning and neuronal health and without it we are at risk of drawing partial and possibly wrong conclusions on how the brain works. 

In the present study we proposed a model that combines functional PCA and Bayesian nonparametric techniques to explore spatio-temporal datasets flexibly. We combined the idea of introducing spatial dependence among curves through the fPC scores proposed by Liu et al.\cite{liu2017functional} with the infinite Gaussian mixture model to obtain a flexible modelling of the covariance structure. The main results show a clear improvement of the PCl-fPCA model both in curve and correlation reconstruction compared to different state-of-the-art fPCA models, particularly in the presence of high noise (as it is often the case in brain recordings) and the ability of exploring curves dependence dynamically allowing for different spatial patterns for each eigendimension retained.

Improvements in the reconstruction of high-noise corrupted curves were also reported by Liu et al.\cite{liu2017functional}; in fact, the beneficial effect of accounting for curves similarity is more evident when the true signal is well masked behind the noise. Nevertheless, a direct modelling of large covariance matrices often resorts to the use of common covariance functions to avoid overparametrisation. The use of functions such as  Mat\'{e}rn or rational quadratic implies a priori knowledge on the shape of spatial dependence. We believe that this approach does not suit highly complex phenomena, such as brain processing, where dependence has a much more elaborate architecture than a simple function of spatial proximity. Clustering the fPC scores allowed us to capture dependence among curve flexibly without the need to estimate the relative spatial covariance matrix. Interestingly, our results suggest that the high flexibility of PCl-fPCA model makes it a very suitable choice even in the cases where a single or even none of the eigendimensions retained support clustering of fPC scores. Further improvements may be derived from modelling the correlation or autocorrelation structure of the noise, although the trade-off with model complexity should be taken into account.\cite{ramsay2005functional}

DP mixture models have also been used for clustering time series through the clustering of the relative coefficients in a basis expansion representation. Many of these works have focused on global clustering, where curves are clustered together \textit{for all} their coefficients.\cite{james2003clustering,ray2006functional,zhou2006bayesian,dunson2008bayesian,bigelow2009bayesian,rodriguez2009bayesian,angelini2012clustering,scarpa2014enriched} However, not only in neuroscientific data, but in many other types of functional data, curves might be characterised by regions of heterogeneous behaviours\cite{petrone2009hybrid}; therefore, some authors have proposed alternative approaches that allow also for local differences in the clustering.\cite{dunson2009nonparametric,maclehose2009nonparametric} In the present study we moved from a global clustering of the data to a local clustering of fPC scores to address both the exploration of brain activity data and to improve curve reconstruction. Dunson et al.\cite{dunson2009nonparametric} and MacLehose et al.\cite{maclehose2009nonparametric} used local clustering only as a means to improve estimation and their methods either neglect inter-subject variability in the coefficients (Dunson et al.\cite{dunson2009nonparametric}) or lack cluster interpretability (MacRose at al.\cite{maclehose2009nonparametric}). 
In contrast, our approach combines the straightforward interpretation of the eigenfunctions with a local clustering of the fPC scores which account for inter-subject variability within each cluster. Therefore, we obtained both an improved curve reconstruction and a rich classification technique. In fact, curves are never identical, they can be potentially assigned to different clusters in each eigendimension, and each eigendimension can have a different number of clusters (see Figure A4 of Supplementary Material WebB for a visual example).
In addition, the assumption of separability of the cross-covariance matrix is avoided and complex time-space interactions are captured by the model; as a consequence, this local borrowing of information also improves the reconstruction of the underlying smooth process. Moreover, we benefit from the properties of the fPCA expansion to tune the hyperparameters and improve the MCMC convergence.

Cross-covariance matrices are often intractable if we do not resort to compromises in our models. A sensible compromise should be tailored to the type of specific data.
In this study, we compromised with the time domain by using fPCA with a fixed number of eigendimensions while giving flexibility in the modelling of spatial dependence. This served the purpose of breaking off from the separability assumption while, at the same time, favouring interpretation and a simple model structure. The fact that the fPCs are treated as known for posterior inference might affect posterior uncertainty. One possible solution to improve coverage is to employ simultaneous credible bounds. These are a finite collection of point-wise intervals, scaled to achieve a specified coverage probability. Existing approaches include those of Besag et al.\cite{besag1995bayesian}; Krivobokova
et al.\cite{krivobokova2010simultaneous} and Crainiceanu et al.\cite{crainiceanu2007spatially} 

By means of a simulation study and the analysis of  fMRI and EEG data, we demonstrate that PCl-fPCA is effective in recovering the underlying smooth curves and it produces a valuable exploration of the spatio-temporal dependence in brain time series. The next step in our approach is the extension to the modelling of multiple subjects' recordings. There are different challenges to consider in the analysis of groups such as the natural inter-individual variability in brain functioning and the dimensionality of the data. We intend to expand our method to replicated data and multiple subjects experiments in our future research.
Exploring inter-individual patterns of functional connectivity and their uncertainty can help answer important questions not only in the study of brain processes but also in the characterisation, early diagnosis and prognosis of brain diseases.

\section*{Acknowledgments}
fMRI data were kindly provided by IRCCS Santa Maria Nascente, Don Gnocchi foundation of Milan (Italy). EEG data were kindly provided by the Henri Begleiter Neurodynamics Laboratory, Department of Psychiatry and Behavioral Sciences, State University of New York (US). We thank the associate editor and reviewers for their useful comments and advice.

\subsection*{Financial disclosure}

None reported.

\subsection*{Conflict of interest}

The authors declare no potential conflict of interests.

\section*{Supporting information}

The following supporting information is available as part of the online article:

\noindent
\textbf{Supplementary material pdf file:}\\
\underline{WebA - Sensitivity Analysis:}
{In the sensitivity analysis we tested how changing the prior expected number of clusters and cluster size impacted on our findings in the application on fMRI data of Section 4 .}

\noindent
\underline{WebB - Additional Figures:}
{Additional figures not shown in the paper that further clarify the features of our model, simulation study and application.}

\noindent
\underline{WebC - Simulation study setting:}
{Additional information on the setting of the simulation study in Section 3.1.}

\noindent
\underline{WebD - Model checks:}
{Additional information on the MCMC checks.}

\section*{Data Availability}
Please note that, upon publication, software in the form of R code will be available from an online repository together with the sample simulated data. EEG data are available at UCI Machine Learning Repository.\cite{EEGData}

\appendix

\section{Posterior conditional distributions\label{app1}}%
In this section we present the posterior conditional distributions for the parameters of our model (Section 2.2).
\begin{align*}
&\xi_{ik}|y_{it},c_{i,k},\mu_{jk},s_{jk},\tau\sim\text{N}\bigg(\frac{\tau\sum_{t=1}^{T}y_{it}\phi_{tk}+s_{jk}\mu_{jk}}{\tau+s_{jk}},\quad\frac{1}{\tau+s_{jk}}\bigg),\\
&\tau|\tilde{\textbf{y}}_{1},...,\tilde{\textbf{y}}_{n},a',b'\sim \Gamma\bigg(\frac{Tn}{2}+a',\quad\dfrac{\sum_{i=1}^{n}\sum_{t=1}^{T}\Big(\tilde{y}_{it}-\sum_{k=1}^{K}\xi_{ik}\phi_{kt}\Big)^{2}}{2}+b'\bigg),\\
&\mu_{jk}|\boldsymbol{c}_{k},\boldsymbol{\xi}_{k},s_{jk},v_{k},r_{k}\sim\text{N}\bigg(\dfrac{s_{jk}\sum_{i:c_{ik}=j}^{n_{jk}}\xi_{ik}+v_{k} r_{k}}{n_{jk}s_{jk}+r_{k}},\quad \frac{1}{n_{jk}s_{jk}+r_{k}}\bigg),\\
&s_{jk}|\boldsymbol{c}_{k},\boldsymbol{\xi}_{k},\beta_{k},z_{k},\mu_{jk}\sim \Gamma\bigg(\frac{n_{jk}}{2} + z_{k},\quad \frac{1}{2}\sum_{i:c_{i}=j}^{n_{jk}}(\xi_{ik}-\mu_{jk})^{2}+\beta_{k}\bigg),\\
&c_{ik}|\boldsymbol{p_{k}},\boldsymbol{\xi_{k}},\boldsymbol{\mu_{k}},\boldsymbol{s_{k}},\alpha_{k}\propto\sum_{j=1}^{J}p_{jk}s_{jk}^{1/2}\exp\Big\{ \dfrac{-s_{jk}}{2}(\xi_{ik}-\mu_{jk})^{2}\Big\},\\
&p_{1k}=\dfrac{p^{'}_{1k}}{\sum_{j=1}^{J}p^{'}_{jk}};\quad p_{jk}=\dfrac{p^{'}_{jk}\prod_{l<j}(1-p_{lk})}{\sum_{j=1}^{J}p^{'}_{jk}},\\
&p^{'}_{jk}|c_{ik},\alpha_{k}\sim\text{Beta}\Big(n_{jk}+1,\quad\alpha_{k}+\sum_{l=j+1}^{J}n_{lk}\Big),\\
&\alpha_{k}|\boldsymbol{p_{k}}\propto\alpha_{k}^{J}\exp\Big\{\alpha_{k}\sum_{j=1}^{J}\log(1-p^{'}_{jk})\Big\};\quad\text{for }\mathcal{S}_{\alpha_{k}}=[0,Q_{k}],
\end{align*}

where $ n_{jk} $ denote the fPC scores in the $ j^{th} $ cluster of the $ k^{th} $ eigendimension and $ \mathcal{S}_{\alpha_{k}} $ the posterior support of $ \alpha_{k} $.

In our model we fixed $a'=b'=10^{-3}$, $ z_{k}=1 $, $ v_{k}=0 $ and the upper-bound for the support of $ \alpha_{k} $  takes into account the dimension-specific features of functional PCA as detailed in the paper, Section 2.2.
\pagebreak
\section*{\Large Supplementary Material}
\section*{Web A - Sensitivity analysis}
In the sensitivity analysis we explored how changing the prior expected number of non-empty clusters and cluster size affected the posterior distribution of the parameters and associated output (Table \ref{TabA}). Overall, our results are robust to changes in the shape and value of the hyperparameters. For the first eigendimension we found only few curves transitioning from cluster 2 to cluster 1 as the prior size and number of clusters increase; vice versa, we found some of the curves with high classification uncertainty (see the analysis of pairwise probabilities in Section 4.2) moving into cluster 2 as the prior size and number of clusters decrease. The other eigendimensions showed only 1 cluster each in every scenario tested.
\begin{table}[H!]
	\caption{fMRI data analysis - 4 different settings tested in the sensitivity analysis of the clusters precision $ s $ (or standard deviation $ \sigma $) and Dirichlet precision $ \alpha $ for each of the $ k= 3 $ dimensions retained.\label{TabA}}\centering
	{\tabcolsep=5.25pt	\begin{tabular} {@{}lcc@{}}
			\hline dimension $ k $ & cluster variability & precision parameter\\
			\hline
			\vspace{4.5pt}	
			\textbf{1}    &  &  \\ 
			\textit{k1} & $ s_{1}\sim\Gamma\big(1, 1.5\times\hat{\lambda}_{1}\big) $ & $ \alpha\sim\text{U}\big[0,15\big] $ \\ 
			\textit{k2} & $ \sigma_{2}\sim\text{U}\big[0, \sqrt{1.5\times\hat{\lambda}_{2}}\big] $ & $ \alpha\sim\text{U}\big[0,10\big] $ \\ 
			\textit{k3} & $ \sigma_{3}\sim\text{U}\big[0, \sqrt{1.5\times\hat{\lambda}_{3}}\big] $ & $ \alpha\sim\text{U}\big[0,10\big] $ \\
			\hline
			\vspace{4.5pt}
			\textbf{2}    &  &    \\ 
			\textit{k1} & $ s_{1}\sim\Gamma\big(1, 0.5\times\hat{\lambda}_{1}\big) $ & $ \alpha\sim\text{U}\big[0,5\big] $  \\ 
			\textit{k2} & $ \sigma_{2}\sim\text{U}\big[0, \sqrt{0.5\times\hat{\lambda}_{2}}\big] $ & $ \alpha\sim\text{U}\big[0,5\big] $ \\ 
			\textit{k3} & $ \sigma_{3}\sim\text{U}\big[0, \sqrt{0.5\times\hat{\lambda}_{3}}\big] $ & $ \alpha\sim\text{U}\big[0,5\big] $ \\
			\hline
			\vspace{4.5pt}
			\textbf{3}    &  &    \\ 
			\textit{k1} & $ s_{1}\sim\Gamma\big(1, \hat{\lambda}_{1}\big) $ & $ \alpha\sim\text{U}\big[0,10\big] $ \\ 
			\textit{k2} & $ s_{2}\sim\Gamma\big(1, \hat{\lambda}_{2}\big) $ & $ \alpha\sim\text{U}\big[0,10\big] $ \\ 
			\textit{k3} & $ s_{3}\sim\Gamma\big(1, \hat{\lambda}_{3}\big) $ & $ \alpha\sim\text{U}\big[0,10\big] $  \\
			\hline
			\vspace{4.5pt}
			\textbf{4}    &  &    \\ 
			\textit{k1} & $ \sigma_{1}\sim\text{U}\big[0, \sqrt{\hat{\lambda}_{1}}\big] $ & $ \alpha\sim\text{U}\big[0,5\big] $ \\ 
			\textit{k2} & $ \sigma_{2}\sim\text{U}\big[0, \sqrt{\hat{\lambda}_{2}}\big] $ & $ \alpha\sim\text{U}\big[0,5\big] $ \\ 
			\textit{k3} & $ \sigma_{3}\sim\text{U}\big[0, \sqrt{\hat{\lambda}_{3}}\big] $ & $ \alpha\sim\text{U}\big[0,5\big] $  \\
			\hline\hline
	\end{tabular}}
\end{table}\\
As the Bayes Factor (BF) in Section 4.2 largely depends on the prior expected number of clusters (see BF formula in Section 2.3), we explored the BF sensitivity to changes in the Dirichlet process precision parameter $ \alpha $. Our results show that the BF for the first eigendimension is consistently $ < 1 $ for all the different priors specified, indicating the probable presence of multiple clusters, while the BFs for the other two eigendimensions are almost always $ \ge 1 $ (Table \ref{TabB}). These indications are in line with results of the other analyses in Section 4.2.
\begin{table}[H!]
	\caption{fMRI data analysis - 6 different settings tested in the sensitivity analysis of the Bayes Factor.\label{TabB}}\centering\vspace{8pt}
	{\tabcolsep=5.25pt	\begin{tabular} {@{}lcc@{}}
			\hline dimension $ k $ &  precision parameter& \textbf{BF}\\
			\hline
			\vspace{4.5pt}	
			\textbf{1}    &  &  \\ 
			\textit{k1} & $ \alpha\sim\text{U}\big[0,10\big] $ & \textbf{0.59} \\ 
			\textit{k2} & $ \alpha\sim\text{U}\big[0,10\big] $ & \textbf{5.07} \\ 
			\textit{k3} & $ \alpha\sim\text{U}\big[0,10\big] $ & \textbf{1.69} \\
			\hline
			\vspace{4.5pt}
			\textbf{2}    &  &    \\ 
			\textit{k1} & $ \alpha\sim\text{U}\big[0,10\big] $  & \textbf{0.45} \\ 
			\textit{k2} & $ \alpha\sim\text{U}\big[0,7\big] $ & \textbf{3.77} \\ 
			\textit{k3} & $ \alpha\sim\text{U}\big[0,7\big] $  & \textbf{2.07}\\
			\hline
			\vspace{4.5pt}
			\textbf{3}    &  &    \\ 
			\textit{k1} &$ \alpha\sim\text{U}\big[0,10\big] $ &\textbf{0.53}  \\ 
			\textit{k2} & $ \alpha\sim\text{U}\big[0,5\big] $ & \textbf{2.93} \\ 
			\textit{k3} & $ \alpha\sim\text{U}\big[0,5\big] $& \textbf{1.33}  \\
			\hline
			\vspace{4.5pt}
			\textbf{4}    &  &    \\ 
			\textit{k1} & $ \alpha\sim\text{U}\big[0,7\big] $ &\textbf{0.53}  \\ 
			\textit{k2} &$ \alpha\sim\text{U}\big[0,7\big] $  & \textbf{4.06}\\ 
			\textit{k3} & $ \alpha\sim\text{U}\big[0,7\big] $ & \textbf{2.31}  \\
			\hline
			\vspace{4.5pt}
			\textbf{5}    &  &    \\ 
			\textit{k1} & $ \alpha\sim\text{U}\big[0,7\big] $ & \textbf{0.44} \\ 
			\textit{k2} &$ \alpha\sim\text{U}\big[0,5\big] $  & \textbf{2.94}\\ 
			\textit{k3} & $ \alpha\sim\text{U}\big[0,5\big] $ & \textbf{0.95}  \\
			\hline
			\vspace{4.5pt}
			\textbf{6}    &  &    \\ 
			\textit{k1} & $ \alpha\sim\text{U}\big[0,5\big] $ & \textbf{0.82} \\ 
			\textit{k2} &$ \alpha\sim\text{U}\big[0,5\big] $  &\textbf{5.53} \\ 
			\textit{k3} & $ \alpha\sim\text{U}\big[0,5\big] $ &\textbf{1.52}   \\
			\hline\hline
	\end{tabular}}
\end{table}

\section*{Web B - Additional figures}
\begin{figure}[H!]
	\centering
\includegraphics[width=0.8\linewidth]{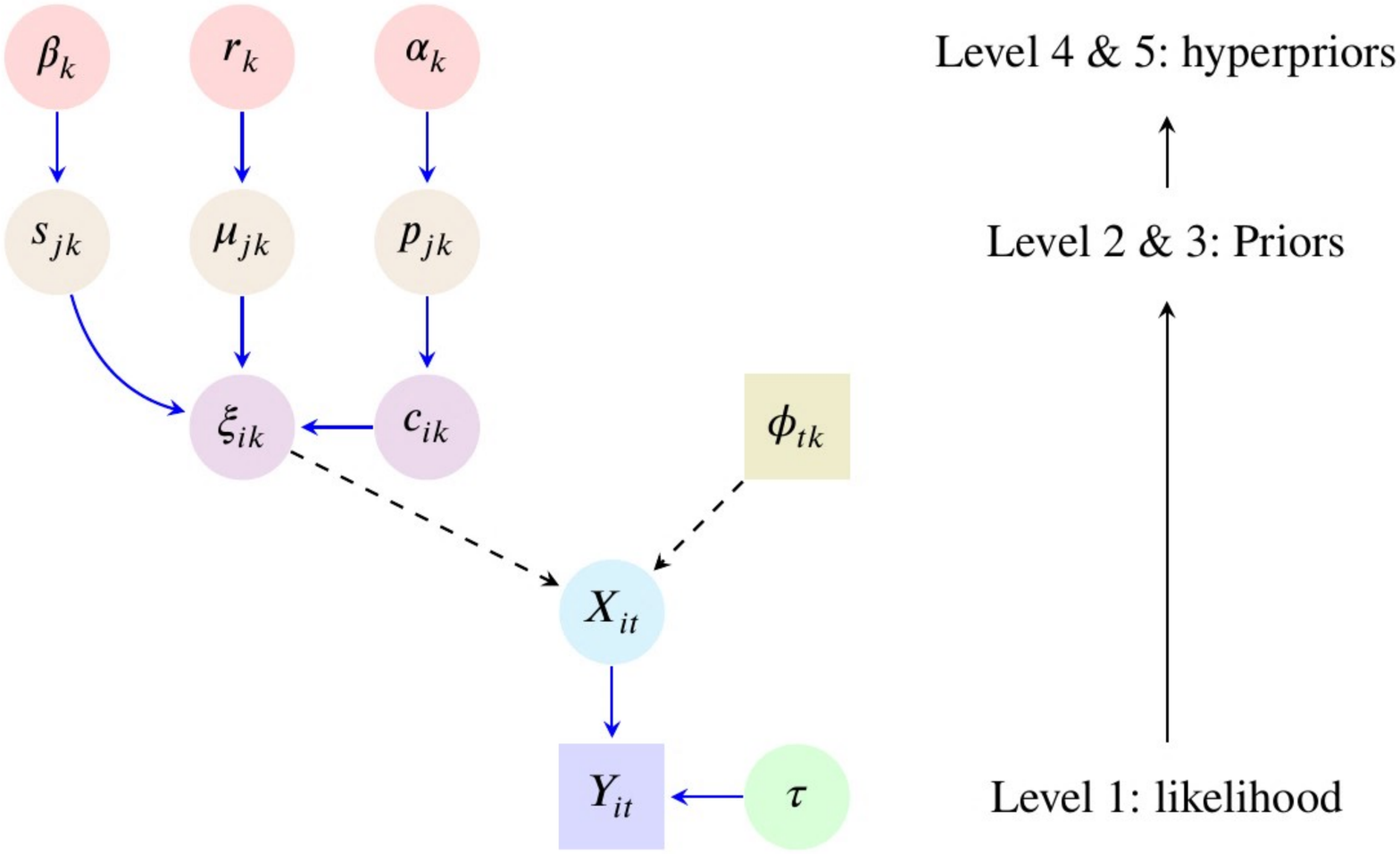}
	\caption{Parameter Clustering fPCA model. Squares = observed random variables; circles =  unobserved random variables; dashed = deterministic relationship; solid = stochastic relationship.\label{B1}}  
\end{figure}

\begin{figure}[H]
	\centering
	\includegraphics[trim={2cm 0 2cm 0},width=1.0\linewidth]{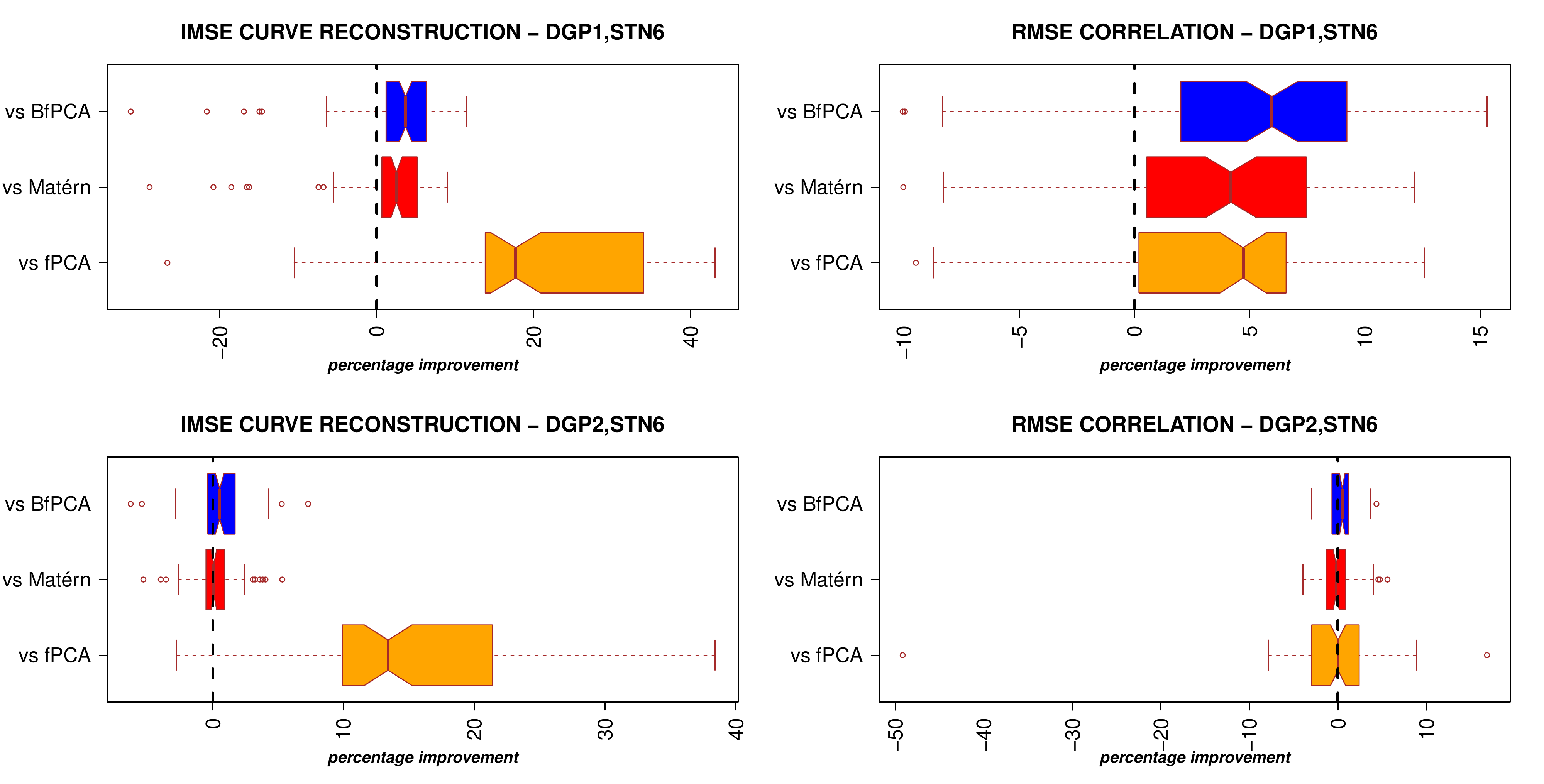}
	\caption{Simulation study: curve and correlation reconstruction for DGP 1 and 2 with low noise (STN6). IMSE and RMSE improvement percentage using PCl-fPCA model versus standard Bayesian fPCA (BfPCA), fPCA model for correlated curves (Mat\'{e}rn) and standard fPCA model (fPCA).\label{B2}}
\end{figure}

\begin{figure}[H]
	\centering
	\includegraphics[trim={2cm 0 2cm 0},width=1.0\linewidth]{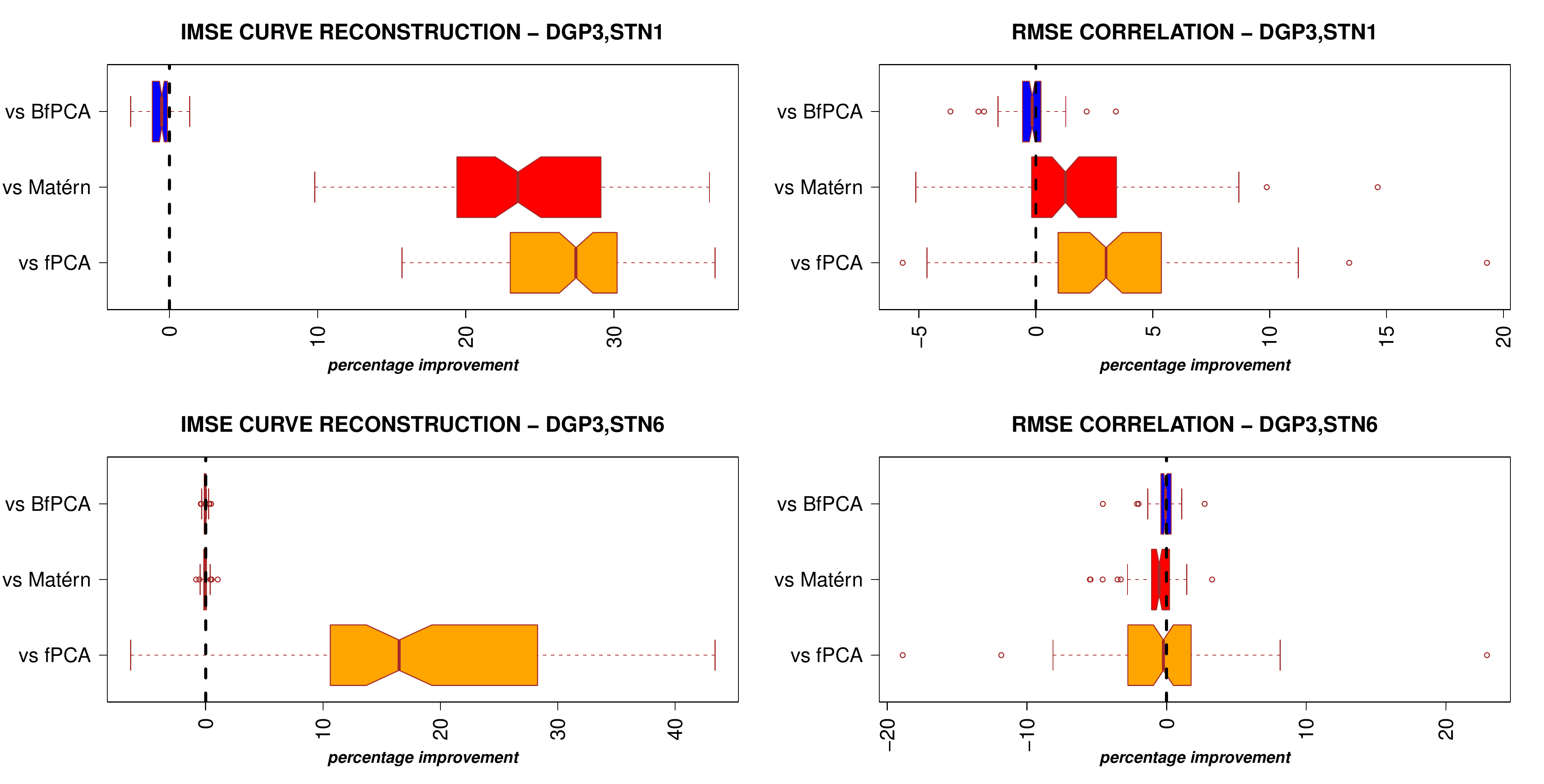}
	\caption{Simulation study: curve and correlation reconstruction for DGP 3 with high and low noise (STN1 and STN6, respectively). IMSE and RMSE improvement percentage using PCl-fPCA model versus standard Bayesian fPCA (BfPCA), fPCA model for correlated curves (Mat\'{e}rn) and standard fPCA model (fPCA).\label{B3}}
\end{figure}

\begin{figure}[H]
	\centering
	\includegraphics[trim={0 3cm 0 3cm},width=0.75\linewidth]{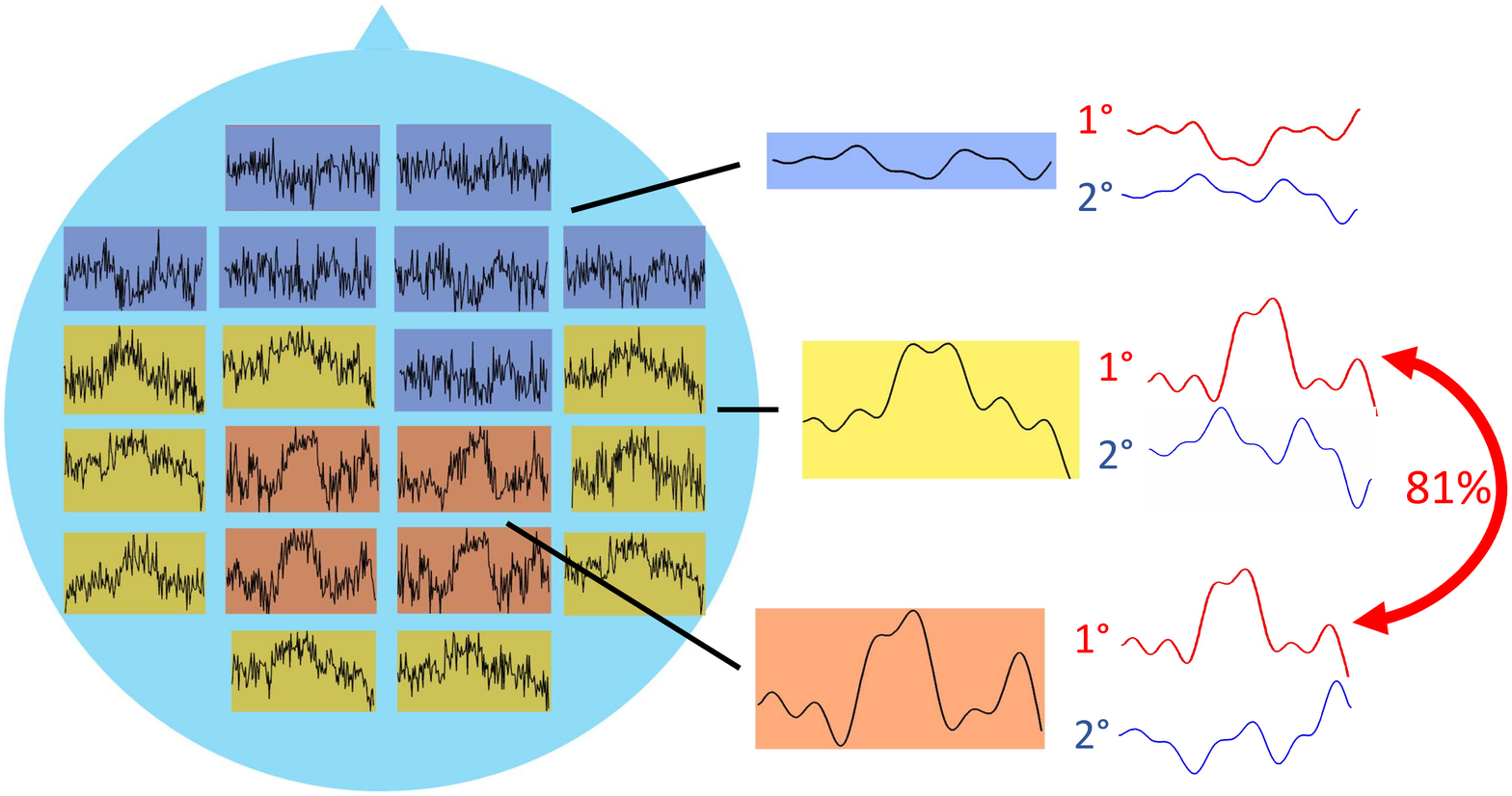}
	\caption{Simulation study - DGP1 scenario: from noise-corrupted time-series to spatio-temporal classified smooth curves. The figure exemplifies information obtained by using PCl-fPCA model. Among 100 noise-corrupted time-series, the model identified three spatial clusters; each group can be represented by the mean of the relative reconstructed curves and the two modes of variation which characterise it (red: 1st eigendimension, blue: 2nd). The new classification method brings to light a strong link (red arrow) between two clusters via the first eigendimension, explaining $ 81\% $ of the variability in the data.}
	\label{B4} 
\end{figure}

\begin{figure}[H]
	\centering
	\includegraphics[width=1\linewidth]{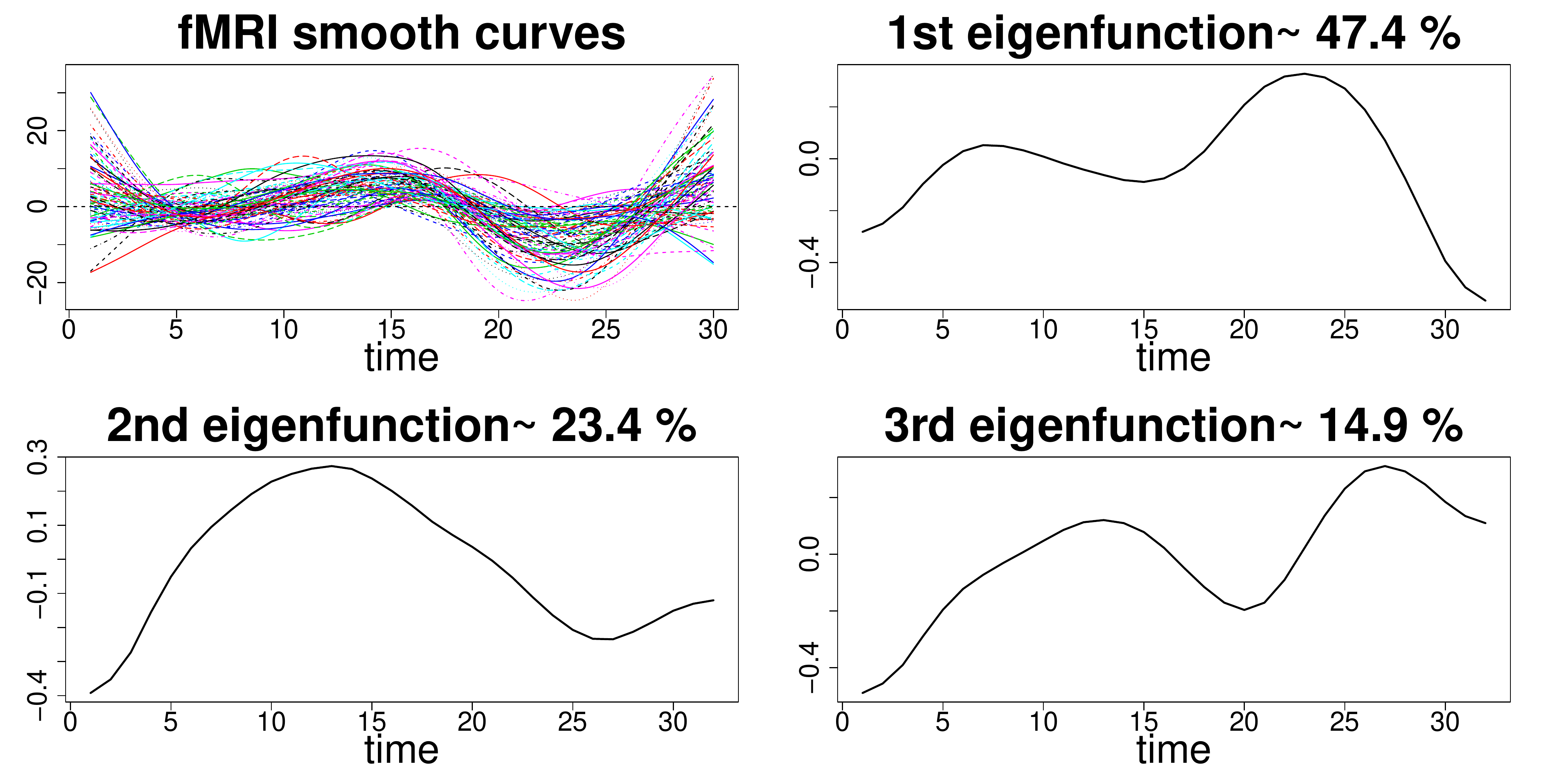}
	\caption{fMRI data analysis: fPCA decomposition. The dataset composed by 90 curves (after smoothing) and the three retained eigendimensions explaining $85\%$ of the total variability in the curves.\label{B5} }	
\end{figure}

\begin{figure}[H]
	\centering
	\includegraphics[trim={10cm 0 7cm 0},width=0.8\linewidth]{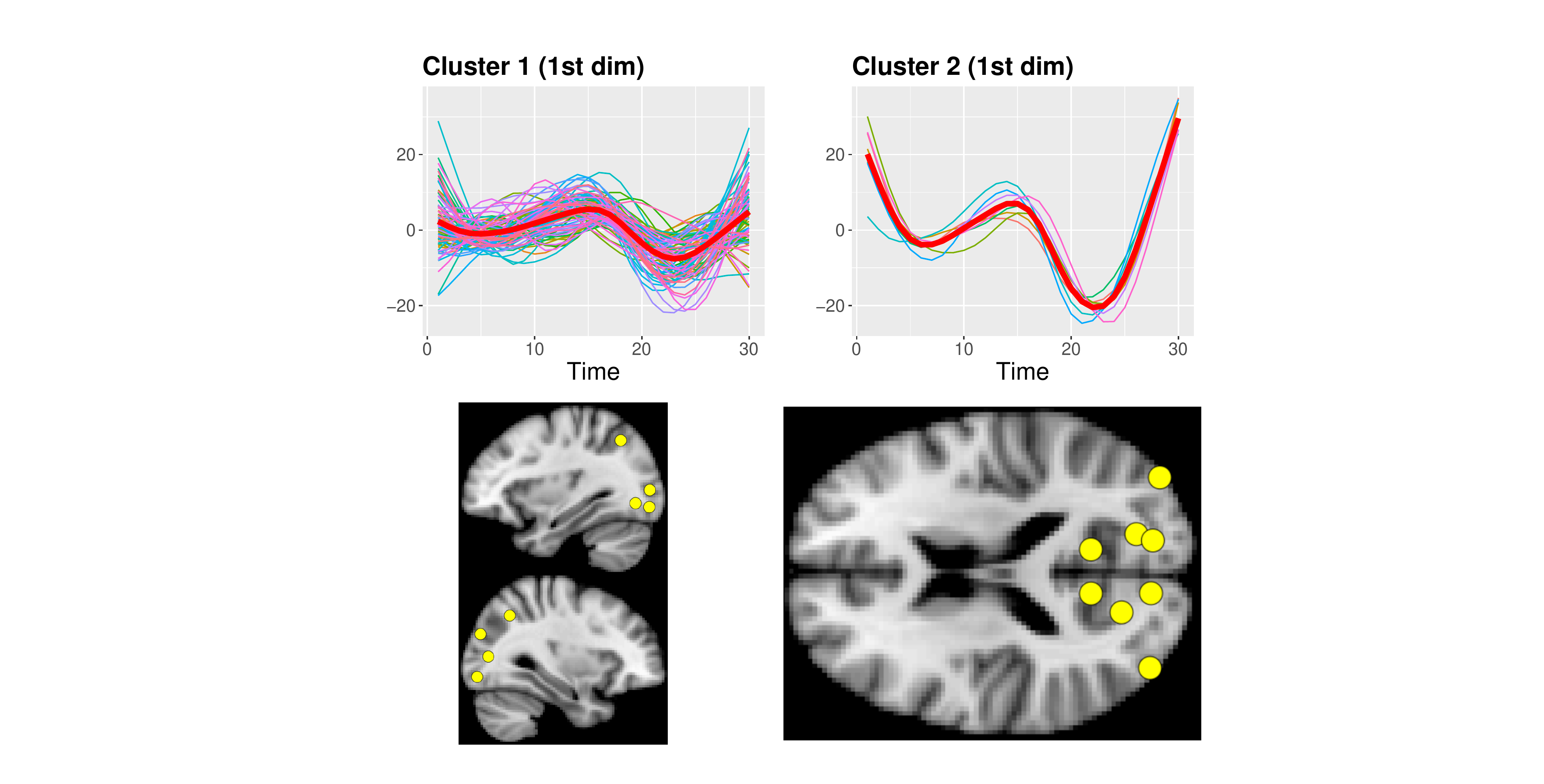}
	\caption{fMRI data analysis: cluster identification with MAPs. Top row: two clusters obtained in the first eigendimension together with the respective overall means (thick line). Bottom row: 3D localisation of cluster 2 (yellow dots) over sagittal and axial slices of human brain.}
	\label{B6} 
\end{figure}

\begin{figure}[H]
	\centering
	\includegraphics[width=0.6\linewidth]{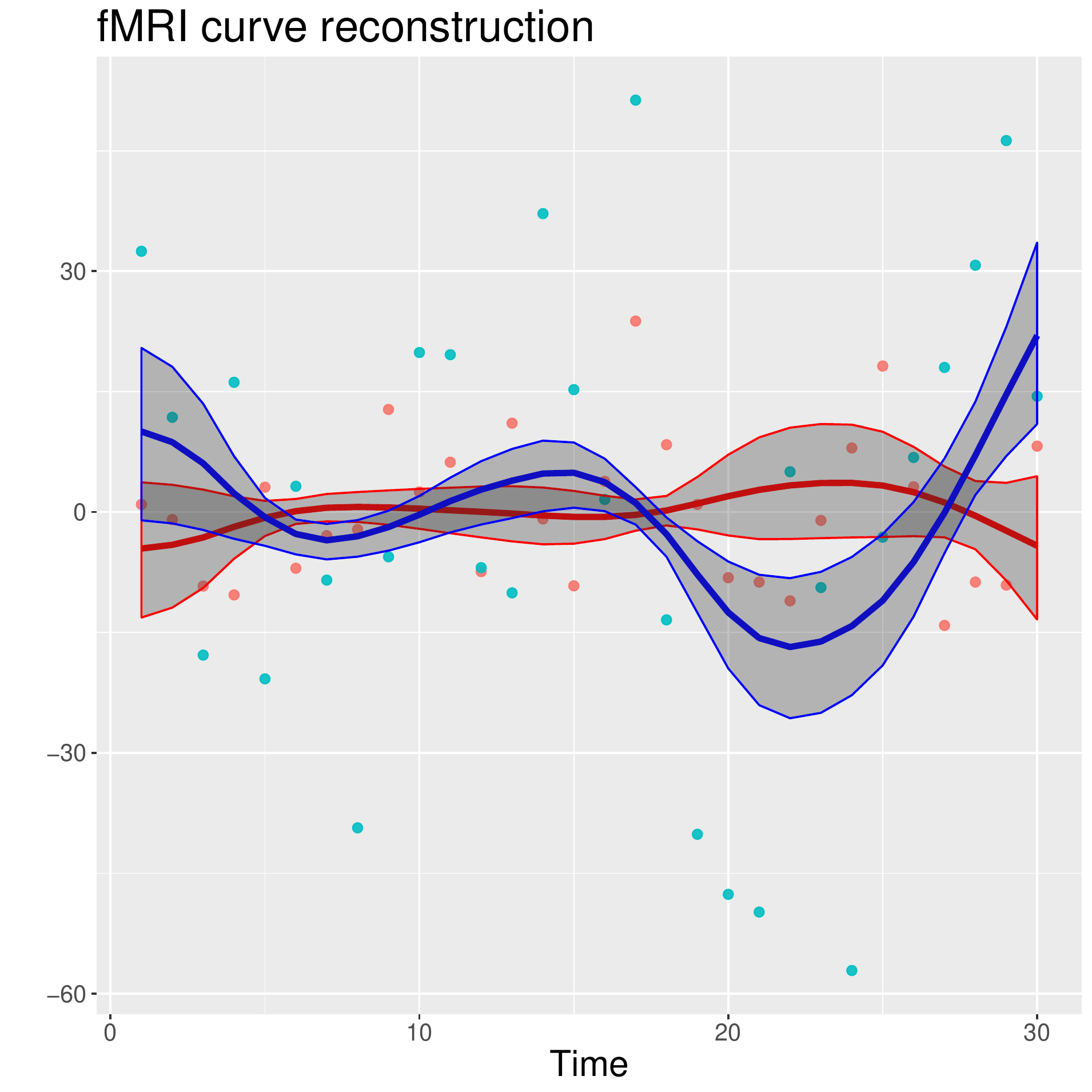}
	\caption{fMRI data analysis: posterior means and pointwise $ 95\% $ credible bands for the resting-state fMRI data recorded from 2 sites in Frontal (red) and Occipital (blue) brain regions.}
	\label{B7} 
\end{figure}

\begin{figure}[H]
	\centering
	\includegraphics[width=0.9\linewidth]{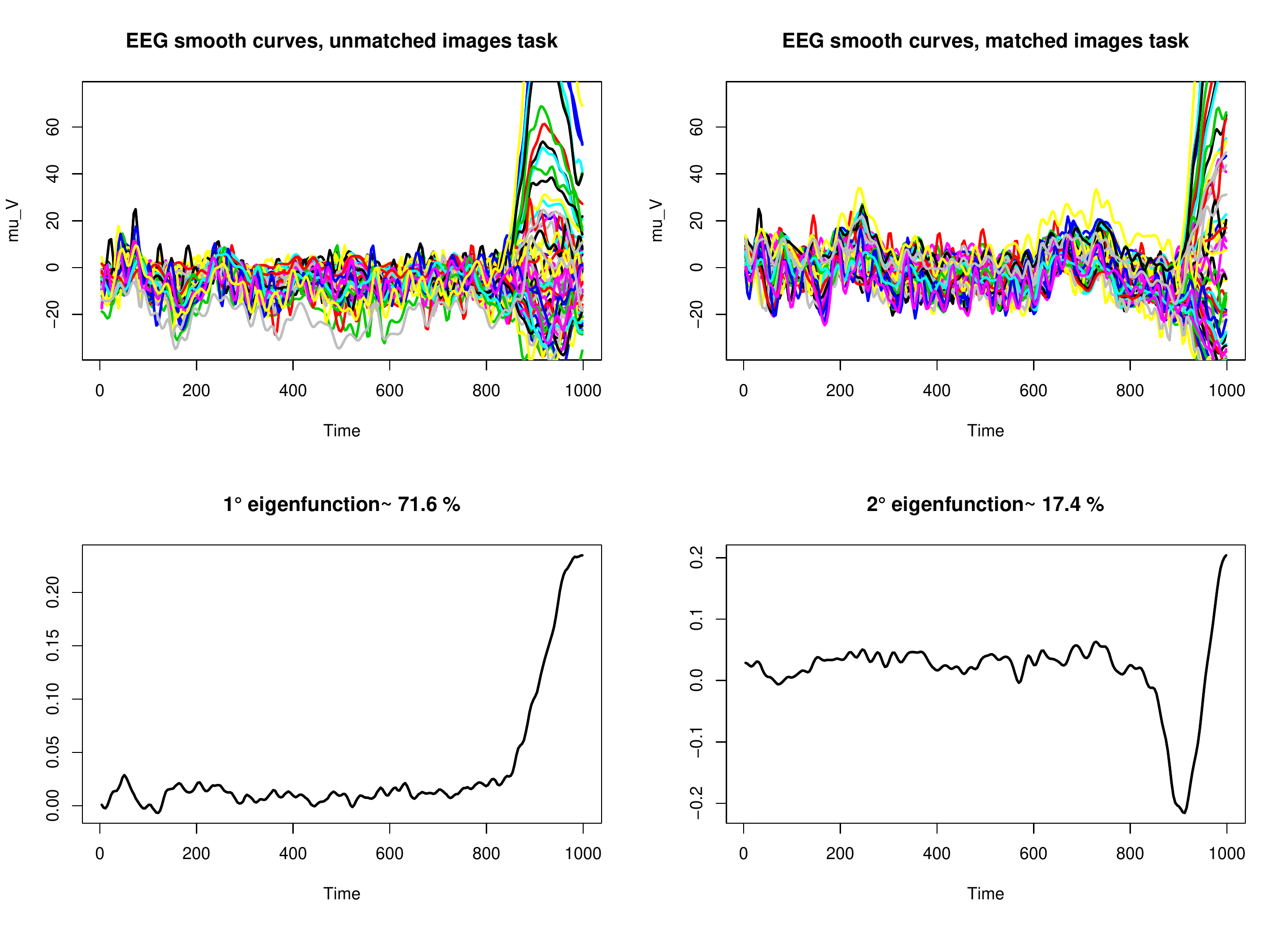}
	\caption{EEG data analysis: fPCA decomposition. The dataset composed by 128 curves (after smoothing) and divided into the two tasks (unmatched and matched images). On the bottom row, the two retained eigendimensions explaining $90\%$ of the total variability in the curves\label{B8}.}
	\label{fig:B8} 
\end{figure}

\begin{figure}[H]
	\centering
	\includegraphics[width=0.95\linewidth,trim={2.2cm 0.2cm 1.9cm 0.3cm},clip]{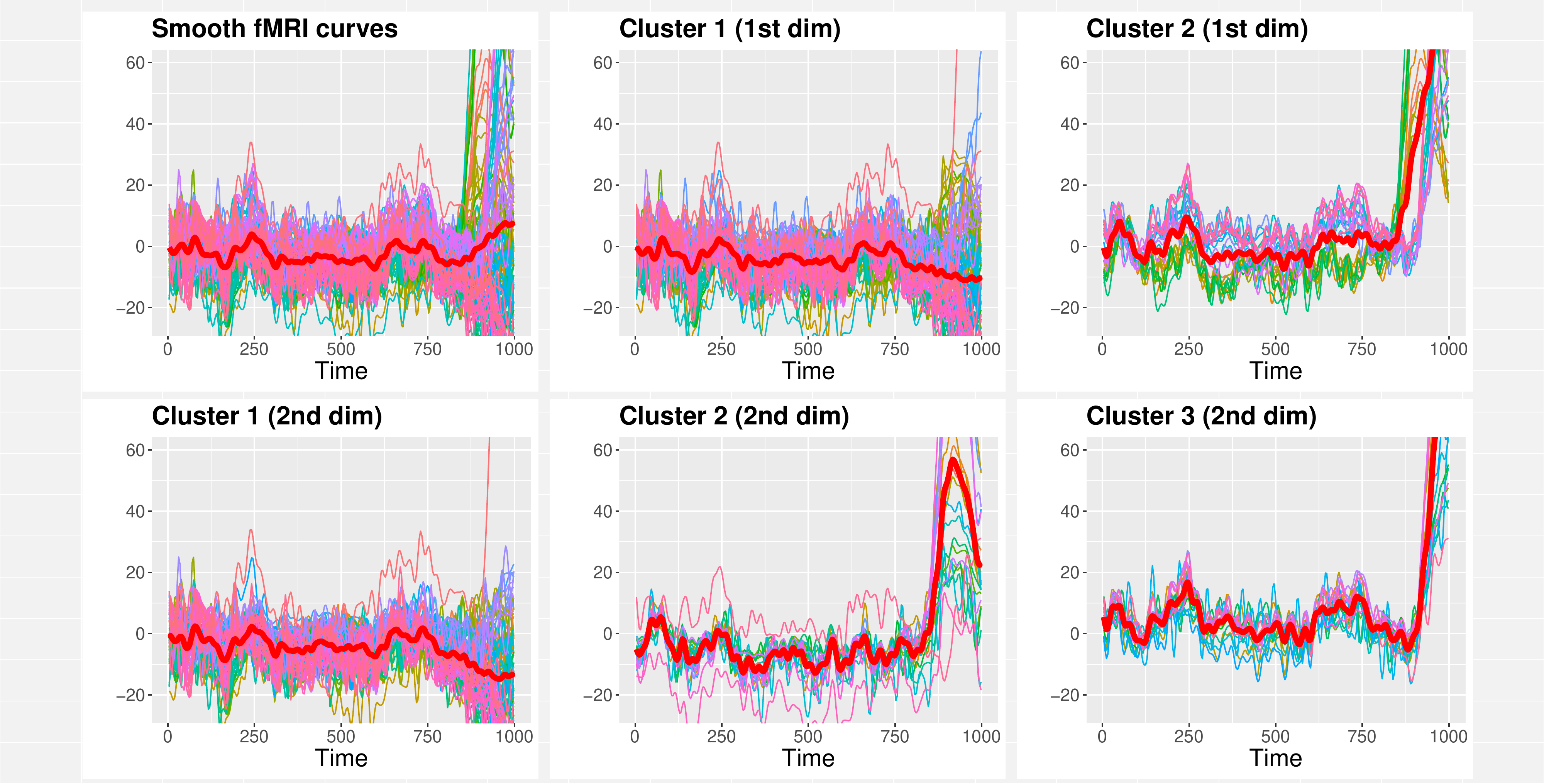}
	\caption{EEG data analysis: curves partition. The 128 EEG curves (after smoothing) are divided into 2 clusters (dimension 1, first row) and 3 clusters (dimension 2, second row) . Cluster-specific means are represented in red\label{B9}.}
	\label{fig:B9} 
\end{figure}

Additional figures not shown in the paper that further clarify the features of our model, simulation study and application.
\begin{itemize}
	\item Figure \ref{B1} shows the Direct Acyclic Graph  representing the structure of the PCl-fPCA model (Section 2.2, main text);
	\item Figures \ref{B2} and \ref{B3} show the result of the simulation study in the case of low noise (STN6) and Data Generating Process 3, where fPC scores correlations were generated by Mat\'{e}rn functions in each eigendimension (Section 3.2, main text);
	\item Figure \ref{B4} exemplifies the results obtained by using PCl-fPCA model (Section 3.2, main text);
	\item Figure \ref{B5} shows the fMRI data used in the first application and the relative eigendimensions retained (Section 4.1, main text);
	\item Figure \ref{B6} provides the results of the application of Maximum a Posteriori Probabilities (MAP) for clustering the fPC scores (Section 4.2, main text);
	\item Figure \ref{B7} shows reconstructed fMRI curves from two different clusters and their credible bands (Section 4.2, main text);
	\item 	Figure \ref{B8} shows the EEG data used in the second application and the relative eigendimensions retained.
	\item Figure \ref{B9} shows the EEG curves partitioned into 2 clusters and 3 clusters for dimension 1 and 2, respectively.
\end{itemize}

\section*{Web C - Simulation study setting}
\subsection*{Data Generating Process 1 (DGP1)} 
Two groups of curves (group 1 and 2) of length $ T=150 $ time points for a total of $ n=100 $ curves were generated by using a mixture of 2 normal distributions on the first eigendimension. 
group 1 (red and orange curves in Figure 1 in the main paper) was composed of $ 50 $ trajectories representing active brain areas, while group 2 (blue curves) represented all other brain regions. Within group 1, two different sub-patterns representing meaningful differences in the way brain areas are activated were obtained by a further partition of the curves into two clusters in the second eigendimension.

\subsection*{Data Generating Process 2 (DGP2)}
The mixture on the first eigendimension in DGP1 was replaced by a zero-centred multivariate Gaussian distribution with covariance matrix defined by a Mat\'{e}rn function $ C_{v}(d) $ in the special exponential case $ v=1/2 $:
\begin{equation*}
C_{1/2}(d)=\sigma^{2}\text{exp}\left\lbrace \frac{-d}{\rho}\right\rbrace ,
\end{equation*}
with $ \rho=0.9 $ and $\sigma^{2}=10$. The second eigendimension was characterised by a mixture of three Gaussian distributions.

\subsection*{Data Generating Process 3 (DGP3)}
The mixtures in the first and second eigendimensions of DGP1 were replaced by multivariate normal distributions with covariance matrix defined by Mat\'{e}rn functions with parameters $  (\rho=0.9,\sigma^{2}=10) $ and $ (\rho=0.7,\sigma^{2}=5) $ respectively.

\subsection*{Computational details and further analyses}
\begin{itemize}
	\item For the Mat\'{e}rn model, we used the approach detailed in Section 3.2 and 3.3 of Liu et al. to capture curve dependence. We employed and adapted the R codes provided by the authors in the on-line supplementary materials. However, we did not employ their cross-covariance local linear smoother and kept the same smoothing approach for all the different models tested to ensure comparability.
	\item We coded the model in \texttt{R} using the \texttt{rjags} package; we employed a conservative approach using $ 100,000 $ iterations for the burn-in and retaining the subsequent $ 100,000 $ MCMC iterations. Convergence diagnostics did not suggest lack of convergence for all parameters of interest. We used a thinning of 5 to store results from 100 simulated datasets efficiently (approximately 70 MB each with K=2). It takes 36 minutes on average to complete one simulation run on a 2-core Intel CPU running at 2.7 GHz with 8 GB RAM.
	\item In order to evaluate the clustering performance of the PCl-fPCA model by taking into account cluster uncertainty we defined, for a given eigendimension $ k $, a measure of distance between the true partition and the pairwise probability matrix which we compared with that of the standard Bayesian fPCA model (i.e. where no cluster is expected). We named the following formula Clustering Improvement Index (CII):
	\begin{align}\label{eq14}
	CII=\dfrac{||M_{\text{std}}-G||_{2}-||M_{\text{new}}-G||_{2}}{||M_{\text{std}}-G||_{2}} \in\bigg[\frac{||M_{\text{std}}-G||_{2}-||\bar{G}-G||_{2}}{||M_{\text{std}}-G||_{2}},1\bigg],
	\end{align}
	where $ M $ represents the clustering matrices for the two competing models (new and standard), $ G $ the underlying truth and $ \bar{G} $ the worst case-scenario we could incur with $ M_{new} $ (i.e. the case where the true 0s and 1s are inverted). It follows that CII is 1 (i.e. $ 100\% $ improvement) when the clustering is completely accurate and values around 0 indicate no real advantage compared with the standard model. Results in Table~\ref{Tab3} are in line with those obtained with ARI highlighting a good clustering performance of the porposed model in both eigendimensions studied for all simulated datasets.
\end{itemize} 	
\begin{table}[H!]
	\caption{Simulation study: clustering performance. The table reports median and IQR of  CII (Clustering Improvement Index) computed for each MC dataset and every STN and eigendimension analysed.\label{Tab3}}\centering
	{\tabcolsep=5.25pt	\begin{tabular} {@{}lc@{}}
			\hline Eigendimension  & $ CII$\\
			\hline
			\vspace{4.5pt}		
			\textbf{STN=1}     &   \\ 	
			\textit{1st dim}  & 0.993 [0.993, 0.986] \\ 
			\vspace{2.5pt}
			\textit{2nd dim}  & 0.645 [0.693, 0.523]\\ 
			\hline
			\vspace{4.5pt}	
			\textbf{STN=6}      &   \\ 	
			\textit{1st dim}   & 0.962 [0.968, 0.950] \\
			\vspace{2.5pt}
			\textit{2nd dim}  & 0.828 [0.855, 0.797]\\ 
			\hline\hline
	\end{tabular}}
\end{table}

\section*{Web D: Applications - model checks} 

In this section we report the results of the MCMC checks for the two applications presented in the manuscript.

\begin{figure}[H!]
	\centering
	\includegraphics[width=0.70\linewidth]{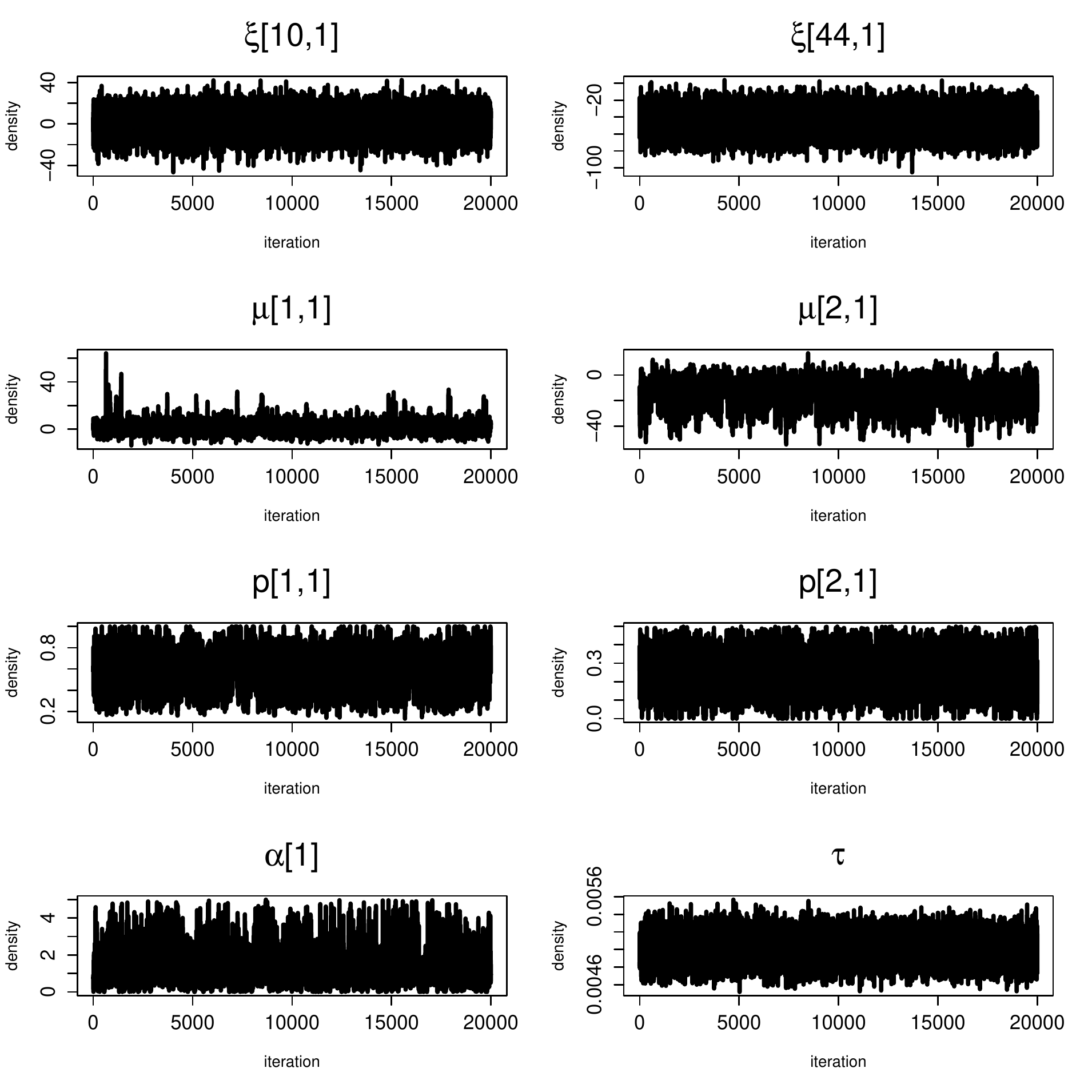}
	\caption{fMRI traceplots (chain 1) for some parameters.\label{B10}}
	\label{fig:B10} 
\end{figure}

\begin{figure}[H!]
	\centering
	\includegraphics[width=0.70\linewidth]{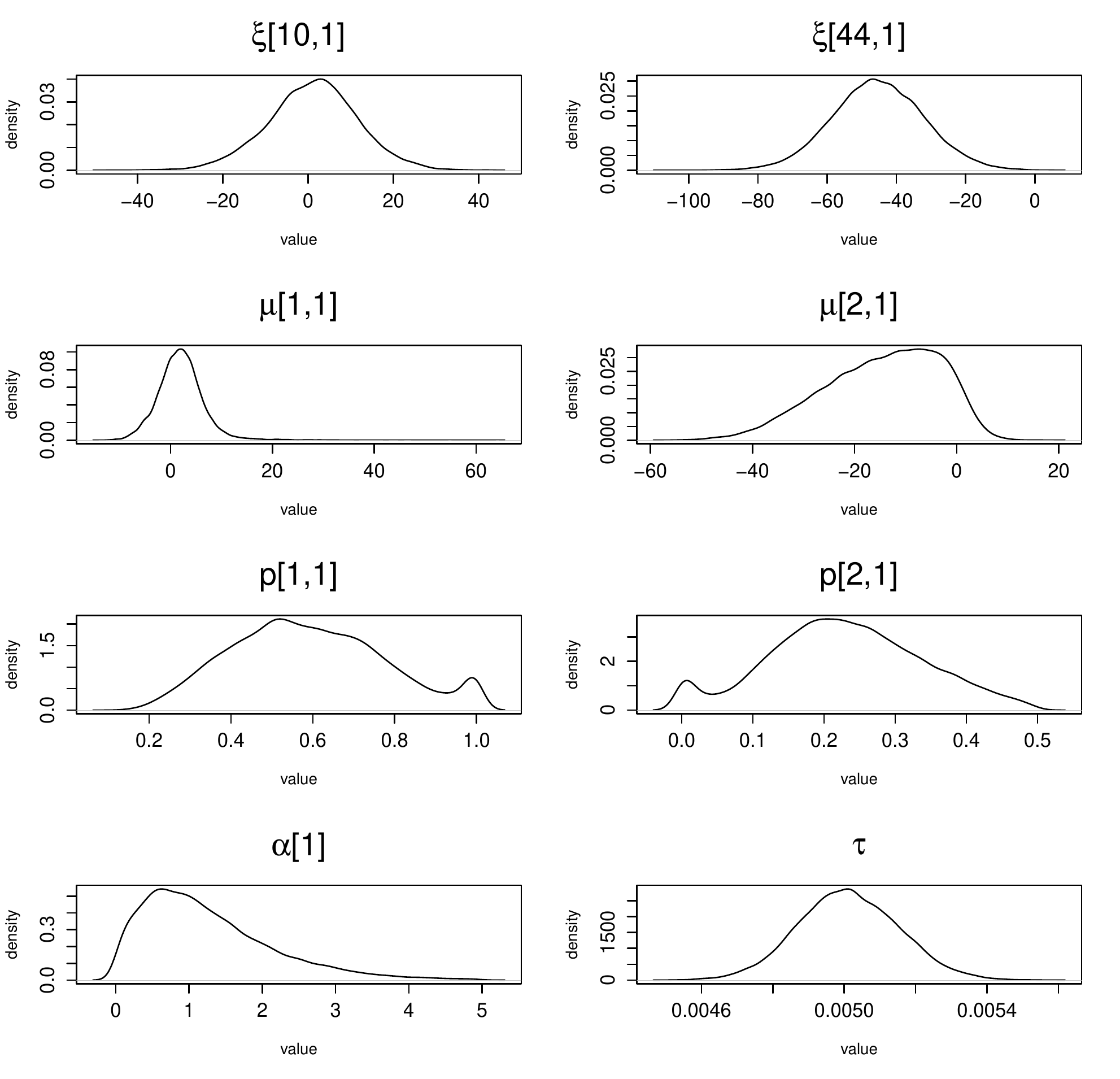}
	\caption{fMRI posterior kernel densities for some parameters.\label{B11}}
	\label{fig:B11} 
\end{figure}

\begin{figure}[H!]
	\centering
	\includegraphics[width=0.70\linewidth]{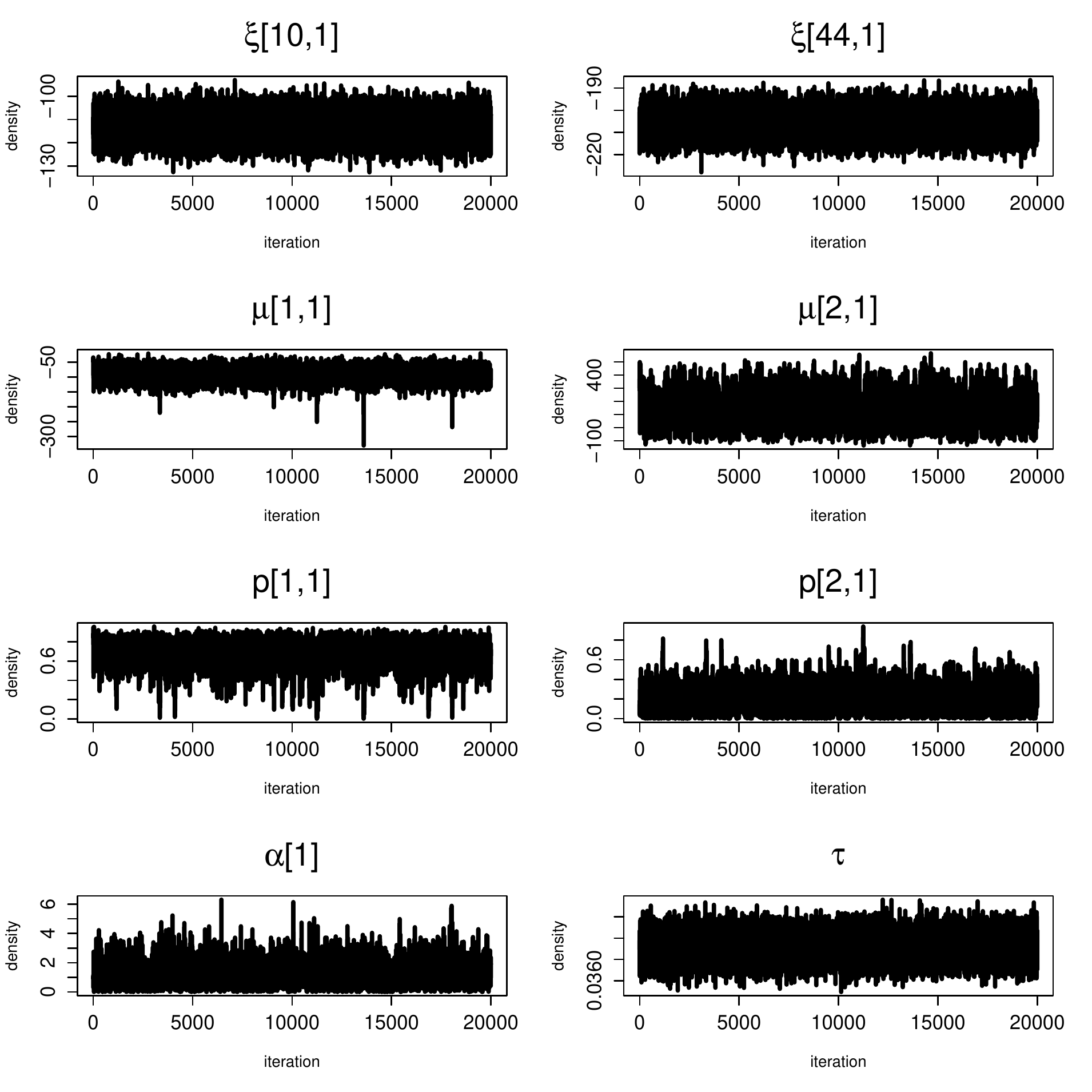}
	\caption{EEG traceplots (chain 1) for some parameters.\label{B12}}
	\label{fig:B12} 
\end{figure}

\begin{figure}[H!]
	\centering
	\includegraphics[width=0.70\linewidth]{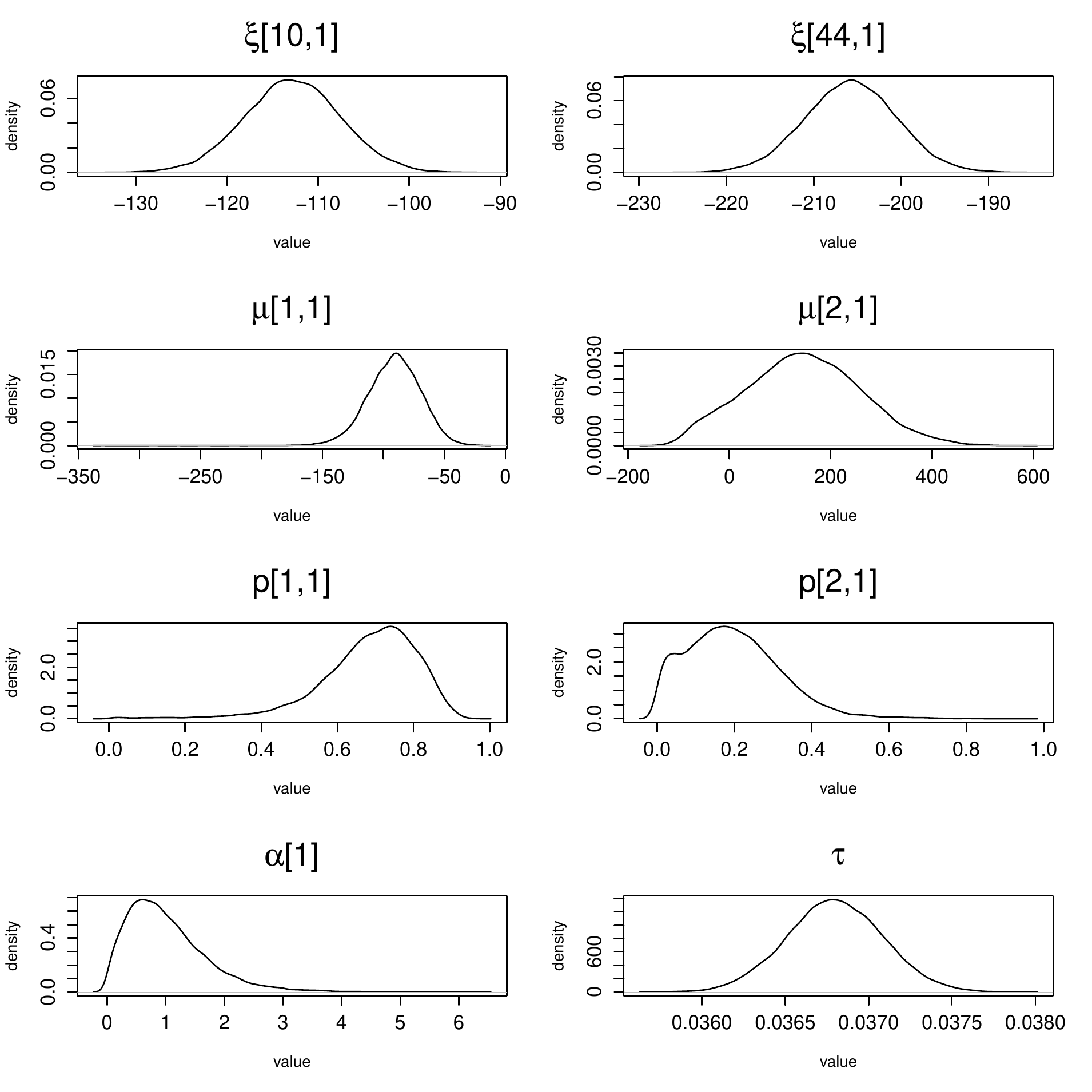}
	\caption{EEG posterior kernel densities  for some parameters.\label{B13}}
	\label{fig:B13} 
\end{figure}

\begin{itemize}
	\item fMRI (664 parameters): BGR statistics (3 chains): mean point estimate = 1.001; $ Q_{0.975} $ upper bound = 1.011. Effective sample size: mean = 6426; $ \text{Q}_{0.025}=518,\,\,\text{Q}_{0.975}=20000$. Traceplots and posterior densities for a selection of important parameters are shown in Figures \ref{B10} and \ref{B11}.
	\item EEG (605 parameters): BGR statistics (3 chains): mean point estimate = 1.008;   $ Q_{0.975} $ upper bound = 1.019 . Effective sample size: mean = 10307; $ \text{Q}_{0.025}=126,\,\,\text{Q}_{0.975}=20520$. Traceplots and posterior densities for a selection of important parameters are shown in Figures \ref{B12} and \ref{B13}.
\end{itemize} 

\bibliographystyle{WileyNJD-v2}
\bibliography{References}%

\end{document}